\documentclass[fleqn,11pt]{article}
\usepackage[cp1251]{inputenc}
\usepackage{amsmath,amssymb}
\usepackage{xcolor}

\newtheorem{stat}{Property}

\newcommand{\Fig}[3]{%
\begin{center}
\parbox{#2cm}{%
\includegraphics[width=#2cm,width=7cm]{#1}}\\
\parbox{8.3cm}{%
\noindent {\refstepcounter{figure} \small \textbf{Fig. \thefigure.}\;
#3}}\end{center}}
\newcommand{\FigH}[4]{%
\begin{flushleft}
\parbox{#2cm}{%
\includegraphics[width=#2cm,height=#3cm]{#1}}
\parbox{8.3cm}{%
\noindent \refstepcounter{figure}\small \textbf{Fig. \thefigure.}\;
#4}\end{flushleft}}
\newcounter{strochka}

\newcounter{spisok}
\usepackage{amsfonts,amssymb,cite}
\usepackage{graphicx}

\def\noi{\noindent}

\newcommand{\Title}[1]{\noi {{\Large\bf #1}}\\[1ex]}

\newcommand{\Author}[2]{\noi{\bf #1}\\[2ex]\noi{\normalsize\it #2}\\}

\newcommand{\Abstract}[1]{\vskip 2mm \begin{center}
        \parbox{16.4cm}{\small\noi #1} \end{center}\medskip}
\newcommand{\foom}[1]{\protect\footnotemark[#1]}

\def\nqq{\hspace*{-2em}}

\usepackage{color}

\def\Jl#1#2{#1 {\bf #2},\ }

\def\ApJ#1 {\Jl{Astroph. J.}{#1}}
\def\CQG#1 {\Jl{Class. Quantum Grav.}{#1}}
\def\DAN#1 {\Jl{Dokl. AN SSSR}{#1}}
\def\GC#1 {\Jl{Grav. Cosmol.}{#1}}
\def\GRG#1 {\Jl{Gen. Rel. Grav.}{#1}}
\def\IJMPD#1 {\Jl{Int. J. Mod. Phys. D}{#1}}
\def\JETF#1 {\Jl{Zh. Eksp. Teor. Fiz.}{#1}}
\def\JETP#1 {\Jl{Sov. Phys. JETP}{#1}}
\def\JHEP#1 {\Jl{JHEP}{#1}}
\def\JMP#1 {\Jl{J. Math. Phys.}{#1}}
\def\NPB#1 {\Jl{Nucl. Phys. B}{#1}}
\def\NP#1 {\Jl{Nucl. Phys.}{#1}}
\def\PLA#1 {\Jl{Phys. Lett. A}{#1}}
\def\PLB#1 {\Jl{Phys. Lett. B}{#1}}
\def\PRD#1 {\Jl{Phys. Rev. D}{#1}}
\def\PRL#1 {\Jl{Phys. Rev. Lett.}{#1}}

\def\lal{&&\nqq {}}

\def\beq{\begin{equation}}
\def\eeq{\end{equation}}
\def\bear{\begin{eqnarray}}
\def\bearr{\begin{eqnarray} \lal}
\def\ear{\end{eqnarray}}
\def\earn{\nonumber \end{eqnarray}}

\begin{document}
\thispagestyle{empty}
\twocolumn[

\vspace{1cm}

\Title{Cosmological models based on an asymmetric scalar doublet with kinetic coupling of components. I. General properties of the cosmological model  \foom 1$^, $\foom 2}

\Author{Yu. G. Ignat'ev$^1$}
    {Institute of Physics, Kazan Federal University, Kremlyovskaya str., 16A, Kazan, 420008, Russia}

 \Author{I. A. Kokh$^2$}
    {N.I. Lobachevsky Institute of Mathematics and Mechanics,  Kazan Federal University, Kremlyovskaya str., 35, Kazan, 420008, Russia}

\Abstract
 {
A mathematical model of the evolution of the Universe, based on an asymmetric doublet of classical and phantom scalar Higgs fields with a kinetic connection between the components, has been constructed and studied. A detailed qualitative analysis was carried out, the properties of the model’s symmetry and invariance with respect to the similarity transformation of fundamental constants were proven. The principles of numerical modeling are formulated and an example of numerical modeling of the evolution of the model for a specific set of fundamental constants and initial conditions is given. The asymptotic behavior of the model near cosmological singularities is studied. It is shown that the cosmological model near singularities manifests itself as an ideal fluid with an extremely rigid equation of state.
 \\[8pt]
 {\bf Keywords}: cosmological model, phantom and classical scalar fields, kinetics interaction, quality analysis, asymptotic behavior, numerical modelling.
}
\bigskip

] 

\section{Introduction}
In \cite{TMF20} a cosmological model based on the \emph{asymmetric scalar Higgs doublet}, in which the classical $\Phi$ and phantom $\varphi$ components interact only through the gravitational field, was studied in detail. Note that earlier in \cite{Vernov1,Vernov2,Vernov3} some one-parameter classes of exact solutions were obtained and studied for a cosmological model based on classical and phantom fields with a special choice of the potential energy function. Later, in the work of \cite{Leon18}, cosmological models based on the scalar quintome with the interaction between the classical and phantom components with a special choice of the potential energy function in the form of an exponential were investigated.  Further, in \cite{Ignat_Sasha_Dima1,Ignat_Sasha_Dima2}, based on a theoretical model of a statistical system of scalar charged fermions with the Higgs interaction potential \cite{TMF_21}, cosmological models were studied in which the interaction between the compo\-nents of an asymmetric scalar doublet was carried out through scalar charged fermions. These works present a rich collection of cosmological models, among which there are models with a finite past (the initial singularity occurs at the final moment of time $t_0>-\infty$) and a finite future (the final singularity occurs at the final moment of time $t_\infty<+\infty$. In particular, in \cite{Ignat_Sasha_Dima2} the behavior of cosmological models near singularities $t\to t_s$ was studied and it was shown that near singularities a total ultrarelativistic equation of state of matter is achieved
\begin{equation}\label{kappa=1/3}
\kappa\equiv \frac{p}{\varepsilon}\to\frac{1}{3},\; (\mbox{at} \ t\to t_s).
\end{equation}

Further in \cite{TMF_24} the transformation properties of cosmological models aimed at transforming the similarities of fundamental constants were studied.

In connection with the problems of the formation of supermassive black holes in the early Universe (see \cite{SMBH1e} -- \cite{Soliton}), in a series of works \cite{Ignatev_SBH} -- \cite{TMF_23} a theory of scalar-gravitational instability of systems was constructed scalarly charged fermions and proposed a model for the formation of supermassive black holes, which satisfactorily explains the main characteristics of these objects. Thus, models of scalar fields with nonlinear interaction may also be important for the astrophysics of the early Universe. In the above series of works, such nonlinear interaction of components was ensured by their interaction with scalar charge carriers. This model, however, has some disadvantages - firstly, extreme mathematical complexity, which makes it difficult to study analytically, and, secondly, insufficiently convincing physical argumentation for the appearance of scalarly charged fermions in the early stages of cosmological evolution.

In this regard, a logical question arises: is it possible to build a cosmological model on a purely field basis, reflecting the main features of a model with scalar charge carriers? In this article we explore such a two-field scalar field model with kinetic interactions between components. In the first part of the article, we will describe the corresponding mathematical model, explore its properties and analyze in detail an example of numerical simulation of a cosmological model with given values of fundamental constants. The second part of the article will be entirely devoted to the results of numerical modeling of cosmological evolution for various values of the fundamental constants of the model and the analysis of these results.

\section{Basic relations of the cosmological model based on an asymmetric scalar doublet}
\subsection{Lagrange function and interaction potential}

Let us consider an asymmetric scalar Higgs doublet (see, for example, \cite{TMF20}) with a kinetic coupling between the components. The corresponding simplest Lagrange function is:
\begin{eqnarray} \label{Lagrange1}
L=\frac{1}{16\pi } (g^{ik} \Phi _{,i} \Phi _{,k} -2V(\Phi ))+\nonumber\\
\!\!\!\frac{1}{16\pi } (-g^{ik} \varphi _{,i} \varphi _{,k} -2v(\varphi ))
+\frac{1}{8\pi}\gamma \biggl(g^{ik}\Phi_{,i}\varphi_{,k}\biggr)^p,
\end{eqnarray}
where
\begin{eqnarray}
\label{V}
V(\Phi )=-\frac{\alpha }{4} \left(\Phi ^{2} -\frac{m^{2} }{\alpha } \right)^{2} ;\\
\label{v}
v(\varphi )=-\frac{\beta }{4} \left(\varphi ^{2} -\frac{\mu^{2} }{\beta } \right)^{2}
\end{eqnarray}
is the potential energy of the corresponding scalar fields, $\alpha$ and $\beta$ are their self-interaction constants, $m$ and $\mu$ are their quanta masses, $\gamma$ is the kinetic interaction constant, $p\geq 1$ is some integer. In contrast to \cite{TMF20}, in the model under consideration, the sign in front of the potential of the classical and phantom fields was specified, corresponding to the real physical model.

By introducing the total potential
\begin{eqnarray} \label{U(F,f)}
\hskip -5pt U(\Phi ,\varphi )=V(\Phi )+v(\varphi )=-\frac{\alpha }{4} \left(\Phi ^{2} -\frac{m^{2} }{\alpha } \right)^{2} \nonumber\\
-\frac{\beta }{4} \left(\varphi ^{2} - \frac{\mu^{2} }{\beta } \right)^{2} \equiv U(\alpha ,\beta ;\Phi ,\varphi ),
\end{eqnarray}
we can draw the following conclusions:

\noindent 1. The potential $U(\Phi ,\varphi )$ has the following symmetries:
\begin{equation} \label{Eq__3_}
U(\pm \Phi ,\pm \varphi )=U(\Phi ,\varphi );
\end{equation}
\begin{equation} \label{Eq__4_}
\hskip -5pt U(-\alpha ,-\beta ;\Phi ,\varphi )=-U(\alpha ,\beta ;\Phi ,\varphi ).
\end{equation}
\noindent 2. The function $U(\Phi ,\varphi )$ has an absolute maximum at the origin of coordinates  ($M_0 (0,0)$) of the phase plane $\{ \Phi ,\varphi \} $.

\noindent 3. For $\alpha >0$, $\beta >0$, the function $U(\Phi ,\varphi )$ has an absolute maximum at points $M_{11} (-m/\sqrt{\alpha } ,-\mu/\sqrt{\beta } )$, $M_{12} (-m/\sqrt{\alpha } ,\mu/\sqrt{\beta } )$, $M_{21} (m/\sqrt{\alpha } ,-\mu/\sqrt{\beta } )$ and $M_{22} (m/\sqrt{\alpha } ,\mu/\sqrt{\beta })$ for $\alpha > 0$ (i.e. $\alpha >0$, $\beta >0$) and the absolute minimum at these points for $\alpha <0$.

Thus, considering the fact that the stationary points of a dynamical system with a Lagrange function of the form \eqref{Lagrange1} coincide with the stationary points of the potential $U(\Phi ,\varphi )$, as well as the parity property of the potential function, we can state the following. Depending on the signs of the parameters $\{\alpha ,\beta \} $ of the potential $U(\alpha ,\beta ;\Phi ,\varphi )$, the corresponding dynamic system can have 2, 6 or 18 stationary points, among which there are attracting ones (absolute minimum), repulsive (absolute maximum) and saddle (conditional extremum) points. \\

\subsection{Reducing the Lagrange function to canonical form}
If we want to construct a theory that is invariant with respect to the signs of the scalar potentials $\Phi,\varphi$, we must choose an even number as $p$. \emph{Further, for simplicity, we will consider the model with $p=1$}.
In this case, the Lagrange function \eqref{Lagrange1} contains an expression that is quadratic with respect to the vectors $\Phi_{,i}$ and $\varphi_{,k}$ (we omit the factor $1/16\pi$):
\begin{equation}\label{Bil_form}
B(\vec{\Phi},\vec{\varphi})= \vec{\Phi}^2-\vec{\varphi}^2+2\gamma(\vec{\Phi}\vec{\varphi}),
\end{equation}
where for simplicity of notation we put $(\vec{a},\vec{b})\equiv g^{ik}a_ib_k$, $\vec{a}^2\equiv g^{ik}a_ia_k$. This form can be reduced to canonical (diagonal) form in the space $\mathbb{R}^2=\{\Phi,\varphi\}$
using linear non-degenerate transformations:
\begin{equation}\label{trans0}
\Phi=a\Phi'+b\varphi';\ \varphi=c\Phi'+d\varphi',
\end{equation}
so that
\[ad-bc\neq0.\]
Substituting the transformation \eqref{trans0} into the expression \eqref{Bil_form}, we get:
\begin{eqnarray}
B(\Phi,\varphi)=\hspace{5.8cm}&\nonumber\\
(a^2-c^2+2ac\gamma)(\vec{\Phi}')^2+(b^2-d^2+2bd\gamma)(\vec{\varphi}')^2&\nonumber\\
+2(ab-cd+ad\gamma+bc\gamma)(\vec{\Phi}'\vec{\varphi}').&\nonumber
\end{eqnarray}
To obtain the canonical (diagonal) form of this form, corresponding to the asymmetric scalar doublet
\begin{equation}\label{B_0}
B_0(\vec{\Phi},\vec{\varphi})= \vec{\Phi}^2-\vec{\varphi}^2,
\end{equation}
we must put:
\begin{eqnarray}\label{alg_sys}
a^2-c^2+2ac\gamma=1;\ b^2-d^2+2bd\gamma=-1;\nonumber\\
ab-cd+ad\gamma+bc\gamma=0.
\end{eqnarray}

Resolving the first two equations of this system for $c,d$, we obtain (the signs are independent):
\begin{eqnarray}\label{c,d}
c=a\gamma\pm \sqrt{a^2(1+\gamma^2)-1};\nonumber\\
d=b\gamma\pm \sqrt{b^2(1+\gamma^2)+1}.
\end{eqnarray}
Substituting these solutions into the last equation of the system \eqref{alg_sys}, we reduce it to the following form:
\begin{equation}\label{Eq_AB}
AB=\pm \sqrt{(A^2-1)(B^2+1)},
\end{equation}
where
\begin{equation}\label{A=a,B=b}
A\equiv a\sqrt{1+\gamma^2};\ B\equiv b\sqrt{1+\gamma^2}.
\end{equation}
By squaring both sides of the equation \eqref{Eq_AB}, we reduce it to the following form:
\begin{equation}\label{A^2_B^2}
A^2-B^2=1.
\end{equation}
The solutions to the equation \eqref{A^2_B^2} are:
\begin{equation}\nonumber
A=\pm\cosh\xi;\; B=\pm\sinh\xi.
\end{equation}

Note that the determinant of the linear trans\-formation matrix \eqref{trans0} is equal to:
\[\frac{\partial(\Phi,\varphi)}{\partial(\Phi',\varphi')}=\pm\frac{1}{\sqrt{1+\gamma^2}}.\]
Negative values of the transform matrix determinant correspond to transformations that do not involve an identity transformation. Choosing the signs in the resulting solutions in such a way that the corresponding transformations also contain the identical transformation, and requiring that at $\gamma=0$ the transformations \eqref{trans0} coincide with the identical one, we finally write down the expressions for the coefficients of this transformation:
\begin{eqnarray}\label{coef_trans0}
a=\frac{1}{\sqrt{1+\gamma^2}}\cosh\gamma; \; b=\pm\frac{\gamma}{\sqrt{1+\gamma^2}}\sinh\gamma;\nonumber\\
c=\pm\sinh\gamma+\frac{\gamma}{\sqrt{1+\gamma^2}}\cosh\gamma;\\
d=\cosh\gamma\pm\frac{\gamma}{\sqrt{1+\gamma^2}}\sinh\gamma.\nonumber
\end{eqnarray}
When passing to these canonical variables, we obtain the canonical form of the kinetic part of the Lagrange function, but at the same time we obtain a much more complicated form of potential energy \eqref{U(F,f)}, which will include, in addition to independent even powers of potentials, products of the form $\Phi \varphi$, $\Phi^2\varphi^2$, $\Phi^3\varphi$, $\Phi\varphi^3$. Due to the linearity of the transformation \eqref{trans0}, which allows us to move to new independent field quantities, it can be argued that the introduction of a kinetic interaction term is equivalent to adding interaction terms to the total potential energy of the system\footnote{The class of such models includes, for example, a model with an exponential potential \cite{Leon18}.}. However, from the point of view of both qualitative theory and numerical modeling, it is easier to preserve the previous field values.

\subsection{Equations of the cosmological model}
Variation of the Lagrange function \eqref{Lagrange1} (see, for example, \cite{Land}) and separation of the second derivatives of the potentials at $p=1$ leads to the field equations:
\begin{eqnarray}
\label{Eq_Phi}
\square \Phi +\frac{V'(\Phi )}{1+\gamma^2}+\frac{\gamma v'(\varphi)}{1+\gamma^2}=0; \\
\label{Eq_varphi}
\square \varphi -\frac{v'(\varphi)}{1+\gamma^2}+\frac{\gamma V'(\Phi )}{1+\gamma^2}=0.
\end{eqnarray}
For the Higgs form of potential energy, these equations take the form:
\begin{equation} \label{Eq_Phi_1}
\square \Phi +m_{*}^{2} \Phi +\gamma \mu_{*}^{2} \varphi=0;
\end{equation}
\begin{equation}
\label{Eq_varphi_1}
\square \varphi +\mu_{*}^{2} \varphi+\gamma m^2_*\Phi =0,
\end{equation}
where $m_{*}$ and $\mu_{*}$ -- effective masses of scalar bosons (pay attention to the signs):
\begin{equation} \label{m_*}
\begin{array}{l}
\displaystyle m_{*} ^{2} =\frac{m^{2} -\alpha \Phi ^{2}}{1+\gamma^2} ; \\
\displaystyle \mu_{*} ^{2} =\frac{-\mu^{2} +\beta \varphi ^{2}}{1+\gamma^2} .
\end{array}
\end{equation}
The energy-momentum tensor of the scalar field with respect to the Lagrange function \eqref{Lagrange1} can be obtained using the standard procedure (see, for example, \cite{Land}):
\begin{eqnarray} \label{T_ik}
T_{ik} =\frac{1}{8\pi }
\biggl(\Phi _{,i} \Phi _{,k} -\frac{1}{2}g_{ik} \Phi _{,j} \Phi ^{,j} +V(\Phi )g_{ik} \biggr)\nonumber\\
+\frac{1}{8\pi } \biggl(-\varphi _{,i} \varphi _{,k} +\frac{1}{2}g_{ik} \varphi _{,j} \varphi ^{,j} +v(\varphi )g_{ik} \biggr)+\nonumber\\
\frac{\gamma}{8\pi}(\varphi_{,i}\Phi_{,k}+\varphi_{,k}\Phi_{,i}-g_{ik}\varphi_{,j}\Phi^{,j}).
\end{eqnarray}

Next, consider the self-consistent system of equations of the cosmological model \eqref{Eq_Phi_1}, \eqref{Eq_varphi_1} and Einstein's equations \eqref{EqEinstein0}\footnote{ We use the Planck system of units: $G=c=\hbar =1$; the Ricci tensor is obtained by contraction of the first and fourth indices $R_{ik}=R^j_{~ikj}$; the metric has signature $(-1,-1,-1,+1)$.}
\begin{equation} \label{EqEinstein0}
R_{ik} -\frac{1}{2} Rg_{ik} =\Lambda_0 g_{ik} +8\pi T_{ik}
\end{equation}
($\Lambda_0$ -- seed value of the cosmological constant), based on a free asymmetric scalar doublet and a spatially flat Friedmann metric (\ref{metric})
\begin{equation}\label{metric}
ds^{2} =dt^{2} -a^{2} (t)(dx^{2} +dy^{2} +dz^{2}),
\end{equation}
assuming
$\Phi =\Phi (t),$ $\varphi =\varphi (t).$ In this case, the energy-momentum tensor \eqref{T_ik} takes on the structure of the energy-momentum tensor of an isotropic liquid
\begin{equation}\label{ten_iso}
T^i_k=(\varepsilon+p)u^iu_k-p\delta^i_k,\qquad(u^i=\delta^i_4)
\end{equation}
with energy density $\varepsilon$ and pressure \textit{p}:
{\small
\begin{eqnarray}
\label{E_sum}
\!\!\! \varepsilon(t)=\varepsilon_{c} +\varepsilon_{f}+ \varepsilon_{cf},\;\; p=p_{c} +p_{f}+p_{cf} ;\hspace{1cm}\nonumber\\
 \label{E_c,E_f}
\!\!\!\!\! \varepsilon_{c} =\frac{1}{8\pi} \biggl(\frac{\dot{\Phi}^{2}}{2} +V(\Phi) \biggr);
\varepsilon_{f} =-\frac{1}{8\pi } \left(\frac{\dot{\varphi}^{2}}{2} -v(\varphi) \right);\nonumber\\
\label{p_c,p_f}
\!\!\!\!\! p_{c} =\frac{1}{8\pi } \left(\frac{\dot{\Phi }^{2}}{2} -V(\Phi) \right);
p_{f} =-\frac{1}{8\pi } \left(\frac{\dot{\varphi }^{2}}{2} +v(\varphi) \right);\nonumber\\
 \label{E_cf-p_cf}
\varepsilon_{cf}=p_{cf}=\frac{\gamma}{8\pi}\dot{\Phi}\dot{\varphi}.\hspace{2cm}
\end{eqnarray}
}
In this case the identity holds:
\begin{equation} \label{Eq__15_}
\varepsilon+p\equiv \frac{1}{8\pi } \dot{\Phi }^{2} -\frac{1}{8\pi } \dot{\varphi }^{2}+\frac{\gamma}{4\pi}\dot{\Phi}\dot{\varphi}.
\end{equation}
So, the system under study consists of two scalar field equations
\begin{eqnarray}\label{old_7a}
\ddot{\Phi }+3\frac{\dot{a}}{a} \dot{\Phi }+\frac{V'_\Phi}{1+\gamma^2}+\frac{\gamma v'_\varphi}{1+\gamma^2} =0,\\
\label{old_7b}
\ddot{\varphi }+3\frac{\dot{a}}{a} \dot{\varphi }-\frac{v'_\varphi}{1+\gamma^2}+\frac{\gamma V'_\Phi}{1+\gamma^2}=0
\end{eqnarray}
and two Einstein equations
\begin{eqnarray} \label{Einstein_44}
^4_4: & 3\frac{\dot{a}^{2} }{a^{2} }-\frac{1}{2}\dot{\Phi}^2+\frac{1}{2}\dot{\varphi}^2-V(\Phi)\nonumber\\
 & -v(\varphi)-\gamma\dot{\Phi}\dot{\varphi}-\Lambda_0=0,\\
\label{Einstein_aa}
^\alpha_\alpha:\ & 2\frac{\ddot{a}}{a}+\frac{\dot{a}^2}{a^2}+\frac{1}{2}\dot{\Phi}^2-\frac{1}{2}\dot{\varphi}^2-V(\Phi)\nonumber\\
 & -v(\varphi)+\gamma\dot{\Phi}\dot{\varphi}-\Lambda_0=0.
\end{eqnarray}

Let us further apply the methodology of \cite{Ignat_Dima_GC20} to obtain a complete system of dynamic equations describing the cosmological model. Let us differentiate Einstein's equation \eqref{Einstein_44} with respect to time and substitute into the resulting equation the values of the second derivatives of scalar potentials from the system of equations \eqref{old_7a} - \eqref{old_7b}. As a result, we get the equation:
\begin{eqnarray}\label{D_Einstein_44}
6H\left(\dot{H}+\frac{1}{2}\dot{\Phi}^2-\frac{1}{2}\dot{\varphi}^2+\gamma\dot{\Phi}\dot{\varphi}\right)=0 \Rightarrow \nonumber\\
\dot{H}=-\frac{1}{2}\dot{\Phi}^2+\frac{1}{2}\dot{\varphi}^2-\gamma\dot{\Phi}\dot{\varphi},
\end{eqnarray}
where
\begin{equation}\label{H}
H=\frac{\dot{a}}{a}
\end{equation}
is the Hubble parameter. Obviously, by subtracting the corresponding parts of the Einstein equation \eqref{Einstein_44} from both sides of the Einstein equation \eqref{Einstein_aa}, we again obtain the equation \eqref{D_Einstein_44}.

Thus, the complete system of equations of our cosmological model is the system \eqref{old_7a}, \eqref{old_7b} and \eqref{D_Einstein_44}. In this case, the Einstein equation \eqref{Einstein_44} is the first zero integral of this system of equations, connecting the initial conditions on the Cauchy hypersurface.

\subsection{Reducing the system of equations to normal form}
Turning to the standard notation of the qualitative theory of dynamical systems (see \cite{Bogoyav})%
\begin{equation}\label{sys_norm}
\frac{dx_{i} }{dt } =F_{i} (x_{1} ,\ldots ,x_{n}) ; {\rm }i=\overline{1,n},
\end{equation}
we write the system of equations \eqref{old_7a} -- \eqref{old_7b}, \eqref{D_Einstein_44} in the form:
\begin{eqnarray}
\label{sys1}
\dot{\Phi}=& \hspace{-3cm} Z &\displaystyle (\equiv F_1);\\
\label{sys2}
\dot{Z}  =& \displaystyle -3HZ-\frac{m^2\Phi -\alpha \Phi ^{3}}{1+\gamma^2}-&\displaystyle\nonumber\\
& \displaystyle \gamma\frac{\mu^{2}\varphi -\beta \varphi ^{3}}{1+\gamma^2}&\displaystyle (\equiv F_2);
\end{eqnarray}
\begin{eqnarray}
\label{sys3}
\dot{\varphi}=&\hspace{-3cm}z,& \displaystyle(\equiv F_3);\\
\label{sys4}
\dot{z} =& \displaystyle -3Hz+\frac{\mu^{2}\varphi -\beta \varphi ^{3}}{1+\gamma^2}- & \displaystyle \nonumber\\
& \displaystyle \gamma \frac{m^2\Phi -\alpha \Phi ^{3}}{1+\gamma^2}& \displaystyle (\equiv F_4);  \\
\label{sys5}
\dot{H} =& \displaystyle-\frac{1}{2}Z^2+\frac{1}{2}z^2-\gamma Zz & \displaystyle(\equiv F_5),
\end{eqnarray}
where $\dot{f}\equiv df{\rm /}dt $.

Let us note that this system is convenient to use when carrying out a numerical analysis of the model, but in a qualitative analysis the choice of this system is unsuccessful, since it leads to its degeneration.. Therefore, when conducting a qualitative study of the system, instead of the equation \eqref{sys5} we will use an equation equivalent to it, which is obtained by adding two Einstein equations \eqref{Einstein_44}, \eqref{Einstein_aa}:
\begin{eqnarray}
\dot{H} =& \displaystyle -3H^2+\frac{1}{2}m^2\Phi^2-\frac{1}{4}\alpha\Phi^4&\nonumber\\
\label{sys6}
& \displaystyle+\frac{1}{2}\mu^2\varphi^2-\frac{1}{4}\beta\varphi^4+\Lambda &\hspace{0cm} (\equiv F^0_5),
\end{eqnarray}
where
\begin{equation}\label{L->L0}
\Lambda=\Lambda_0-\frac{m^4}{4\alpha}-\frac{\mu^4}{4\beta}
\end{equation}
is the observed value of the cosmological constant.
The system of equations \eqref{sys1} -- \eqref{sys5} represents the mathematical model of the dynamic system under study
in 5-dimensional phase space
\begin{equation}\label{R5}
\mathbb{R}^5 = \{\Phi,Z,\varphi,z, H\}\equiv$ \\ $\{x_1,x_2$,$x_3,x_4,x_5\}.
\end{equation}
Let us introduce normalized quantities that will be useful later - \emph{normalized effective energy, $E$, and normalized effective pressure, $P$, of the system}:
\begin{eqnarray}\label{E_m}
E=& 8\pi \varepsilon(t)+\Lambda_0\equiv \displaystyle\frac{Z^2}{2}-\frac{\alpha\Phi^4}{4}+\frac{m^2\Phi^2}{2}\nonumber\\
&\displaystyle-\frac{z^2}{2}-\frac{\beta\varphi^4}{4}+\frac{\mu^2\varphi^2}{2}
+\gamma Zz+\Lambda,
\end{eqnarray}
\begin{eqnarray}\label{P_m}
P=& 8\pi p(t)-\Lambda_0\equiv \displaystyle\frac{Z^2}{2}+\frac{\alpha\Phi^4}{4}-\frac{m^2\Phi^2}{2}\nonumber\\
&\displaystyle-\frac{z^2}{2}+\frac{\beta\varphi^4}{4}-\frac{\mu^2\varphi^2}{2}
+\gamma Zz-\Lambda.
\end{eqnarray}
Then the equation \eqref{Einstein_44} can be written as follows:
\begin{equation}\label{3H^2}
3H^2-E=0.
\end{equation}
This equation defines an Einstein hypersurface in phase space, the explicit form of which is:
\begin{equation}\label{Hip_Ein}
\begin{array}{l}
3H^2-\dfrac{Z^2}{2}+\dfrac{\alpha\Phi^4}{4}-\dfrac{m^2\Phi^2}{2}\\[11pt]
+\dfrac{z^2}{2}+\dfrac{\beta\varphi^4}{4}-\dfrac{\mu^2\varphi^2}{2}
-\gamma Zz-\Lambda=0.
\end{array}
\end{equation}

Note that due to the non-negativity of $H^2$, an important property follows from \eqref{3H^2}:
\begin{equation}\label{E>0}
E\geq0.
\end{equation}

In this case, the first integral of the dynamic system \eqref{sys1} -- \eqref{sys5} (Einstein's equation \eqref{Einstein_44}) can be written in the form (see \cite{TMF20})
\begin{equation}\label{3H^2c}
3H^2-E=\rm{const}.
\end{equation}
The Einstein equation corresponds to the zero value of this integral, therefore the Hubble parameter $H$ can be algebraically expressed from the equation \eqref{3H^2} through the remaining dynamic variables and the resulting relation can be used to determine its initial value.

So, the cosmological model under study can be considered as a five-dimensional dynamical system in the arithmetic phase space $\mathbb{R}^5=\{\Phi,Z,\varphi,z,H\}=\mathbb{R}^3 \cup \mathbb{R}^3$, three-dimensional phase
 subspaces $\mathbb{R}^3=\{\Phi,Z,H\}$ and $\mathbb{R}^3=\{\varphi,z,H\}$ will be further denoted by the symbols $\Sigma_{\Phi}$ and $\Sigma_{\varphi}$ and, for simplicity, called the classical and phantom phase spaces, respectively. In this case -- $\Sigma_{\Phi}\cup\Sigma_{\varphi}=\mathbb{R}^5$ and $\Sigma_{\Phi}\cap\Sigma_{\varphi}=\mathbb{R}^1=OH$.

Using effective energy and pressure, the effective coefficient {\it barotrope} of cosmological matter $\kappa$ \eqref{kappa=1/3} is determined,
which, in turn, determines the invariant cosmological acceleration
\begin{equation}\label{Omega}
\Omega=\frac{a\ddot{a}}{{\dot{a}}^2}\equiv 1+\frac{\dot{H}}{H^2}=-\frac{1}{2}(1+3\kappa).
\end{equation}

\subsection{Dynamic system with scale factor}
The system of dynamic equations under study \eqref{sys1}--\eqref{sys5} for variables in the 5-dimensional phase space $\mathbb{R}^5 = \{\Phi,Z,\varphi,z, H\}$ is complete and quite sufficient for solving most problems of cosmology. However, in cases where it is necessary to study the behavior of the scale factor near the cosmological singularity, this system is not very convenient, since it requires the calculation of definite integrals near the singularity. Therefore, in these cases it is more convenient to consider a 6-dimensional dynamical system in the space $\mathbb{R}^6 = \{\xi,\Phi,Z,\varphi,z, H\}$:
\begin{eqnarray}
\label{sys0a}
\dot{\xi}= H \qquad (a\equiv \exp(\xi));\\
\label{sys1a}
\dot{\Phi}=  Z; \\
\label{sys2a}
\dot{Z}  =  -3HZ-\frac{m^2\Phi -\alpha \Phi ^{3}}{1+\gamma^2}-\gamma\frac{\mu^{2}\varphi -\beta \varphi ^{3}}{1+\gamma^2};\\
\label{sys3a}
\dot{\varphi}=z;\\
\label{sys4a}
\dot{z} = -3Hz+\frac{\mu^{2}\varphi -\beta \varphi ^{3}}{1+\gamma^2}- \gamma \frac{m^2\Phi -\alpha \Phi ^{3}}{1+\gamma^2};  \\
\label{sys5a}
\dot{H} = -\frac{1}{2}Z^2+\frac{1}{2}z^2-\gamma Zz.
\end{eqnarray}
Note that the equation \eqref{sys0a} is actually the definition of $H(t)$.
\subsection{Symmetries of a dynamical system}
A particular cosmological model $\textbf{M}$ is determined, firstly, by a system of dynamic equations \eqref{sys1}--\eqref{sys5}, secondly, by fundamental parameters
\begin{equation}\label{P}
\textbf{P} = [[\alpha,\beta, m, \mu],\gamma,\Lambda]
\end{equation}
and, thirdly, by initial conditions
\begin{equation}\label{I}
\textbf{I} = [ \Phi_0,Z_0,\varphi_0, z_0,e],\qquad (e=\pm 1).
\end{equation}
Here we took into account that the initial value of the Hubble parameter $H_0$, up to sign, can be obtained from the first integral \eqref{3H^2} taking into account the definition of the energy density \eqref{E_m}. The sign of $H_0$ is taken into account by the indicator $e$ in the formula \eqref{I}.

In \cite{TMF20} the symmetry properties of the dynamical system \eqref{sys1} -- \eqref{sys5} were studied in relation to the reflection transformation of the basic functions of the model and the time variable at $\gamma\equiv0$. All these invariance properties are preserved in our model.

Let us find out how the solutions of the system \eqref{sys1} -- \eqref{sys5} depend on the sign of the interaction constant $\gamma$. Substituting $\gamma\to-\gamma$ into the equations \eqref{sys1} -- \eqref{sys5}, we arrive at
to the next conclusion.
\begin{stat}\label{stat1}
The dynamic system \eqref{sys1} -- \eqref{sys5} is invariant under the following transformations
\begin{eqnarray}\label{t->t}
\gamma\to -\gamma;\; t\to t;\;  H\to H; \\
\left\{\begin{array}{l}
\Phi\to-\Phi,Z\to-Z,\varphi\to\varphi,z\to z;\\
\Phi\to\Phi, Z\to Z,\varphi\to-\varphi,z\to-z;\\
\end{array}\right.\nonumber
\end{eqnarray}
\begin{eqnarray}\label{t->-t}
\gamma\to -\gamma;\; t\to -t;\;  H\to -H; \\
\left\{\begin{array}{l}
\Phi\to-\Phi,Z\to Z,\varphi\to\varphi,z\to -z;\\
\Phi\to\Phi, Z\to -Z,\varphi\to-\varphi,z\to z;\\
\end{array}\right.\nonumber
\end{eqnarray}
In this case, of course, the corresponding transformations must be made in the initial conditions \eqref{I}.
\end{stat}

%
\subsection{Scale transformations of the cosmological model}
By analogy with \cite{TMF_24} we formulate the following statement.
\begin{stat}\label{stat2}
Let the solution to the Cauchy problem for the model $\textbf{M}$ be:
\begin{equation}
\textbf{S}(t)= [ \Phi(t), Z(t),\varphi(t), z(t), H(t)].
\end{equation}
In what follows we will call this model \emph{preimage}.
Let us consider a similar model ${\bf\tilde{M}}$, which we will further call \emph{image} and which is obtained from the inverse image of $\textbf{M}$ \emph{similarity transformation} with coefficient $k$
:
\begin{eqnarray}\label{sim1}
{\bf\tilde{P}} = \left[\left[\frac{\alpha}{k^2},\frac{\beta}{k^2}, \frac{m}{k}, \frac{\mu}{k}\right],\gamma,\frac{\Lambda}{k^2}\right];\\
{\bf \tilde{I}} = \left[ \Phi_0,\frac{Z_0}{k},\varphi_0 ,\frac{z_0}{k},e\right].
\end{eqnarray}
Then the time $\tilde{t}$ in the model ${\bf\tilde{M}}$ is related to the inverse image of $t$ by the relation:
\begin{equation}\label{sim2}
\tilde{t}=kt,
\end{equation}
and the solution to the corresponding Cauchy problem is related as follows
\begin{equation}\label{sim3}
\!\!\!\!\begin{array}{l}
\textbf{S}(t)= [ \Phi(t),  Z(t), \varphi(t), z(t), H(t)] \Rightarrow\\
{\bf \tilde{S}}(t)= \left[ \Phi\left(\frac{t}{k}\right), \frac{1}{k}Z\left(\frac{t}{k}\right), \varphi\left(\frac{t}{k}\right), \frac{1}{k}z\left(\frac{t}{k}\right),\frac{1}{k} H\left(\frac{t}{k}\right)\right].
\end{array}
\end{equation}
Note that the extended system \eqref{sys0a}--\eqref{sys5a} is also invariant under similarity transformations.
\end{stat}

%

\subsection{Real ranges and movement near energetic hypersurfaces}

In order for the system of differential equations \eqref{sys1} -- \eqref{sys5} to have a real solution, the effective energy  $E$ of the system \eqref{E>0} must be non-negative:
\begin{equation}\label{E>=0_2}
\begin{array}{ll}
&\displaystyle
\frac{Z^2}{2}-\frac{\alpha\Phi^4}{4}+\frac{m^2\Phi^2}{2}-\vspace{5pt}\\
-&\displaystyle\frac{z^2}{2}-\frac{\beta\varphi^4}{4}+\frac{\mu^2\varphi^2}{2}
+\gamma Zz+\Lambda\ge 0.
\end{array}
\end{equation}

A unique feature of the system under consideration is a change in the topology of the phase space due to the appearance in it of areas in which movement is impossible. These regions are distinguished by the condition that the effective total energy \eqref{E_m} is negative, while the regions accessible to phase trajectories are determined by the condition (\ref{E>=0_2}). Hypersurfaces of zero effective energy $S_{3}^{0} \subset \mathbb{R}^{4}$, dividing the phase space into regions of permissible and forbidden values of dynamic variables, are described by algebraic equations of the 4th degree with respect to phase variables:
\begin{eqnarray} \label{Eqs12_}
E(\Phi,Z,\varphi,z) =0\Rightarrow \frac{Z^2}{2}-\frac{\alpha\Phi^4}{4}+\frac{m^2\Phi^2}{2}\nonumber\\
-\frac{z^2}{2}-\frac{\beta\varphi^4}{4}+\frac{\mu^2\varphi^2}{2}+\gamma Zz+\Lambda= 0.
\end{eqnarray}

Let us note \label{ZamGamma} a circumstance that qualitatively distinguishes the cosmological model with an asymmetric scalar doublet from the corresponding model with a single scalar field. The phase space hypersurface $S_{3}^{0} \subset \mathbb{R}^4$ \eqref{Eqs12_} is determined only by the parameters $\textbf{P}$ \eqref{P} of the field model of the scalar doublet   and does not depend on the time variable $t$.
However, the intersection of two-dimensional phase planes of single fields $\Pi_\Phi=\{\Phi,Z\}$ and $\Pi_\varphi=\{\varphi,z\}$ with the hypersurface $S^0_3$ can be two-dimensional curves $\Gamma_\Phi$ and $\Gamma_\varphi$ (closed or open), essentially depending on the values of the dynamic variables of another scalar field, and therefore depending on the temporary variable:
\begin{eqnarray}\label{Gamma}
\Gamma_\Phi(t): \; E(\Phi,Z,t)=0\Rightarrow E(\Phi,Z,\varphi(t),z(t)) =0;\nonumber\\
\Gamma_\varphi(t): \;E(\varphi,z,t)=0\Rightarrow E(\Phi(t),Z(t),\varphi,z) =0.\nonumber
\end{eqnarray}
As a result, the topology of the two-dimensional phase subspaces $\Pi_\Phi$ and $\Pi_\varphi$ can change significantly with time.

\section{Qualitative analysis of the cosmological model}
\subsection{Singular points of a dynamical system \label{ST}}
First of all, we note that the singular points of the extended dynamical system \eqref{sys0a}--\eqref{sys5a} at finite cosmological times must coincide with the singular points of the system \eqref{sys1}--\eqref{sys5} (see \cite{ Similarity}). This allows us to limit ourselves to a qualitative analysis of the main dynamic system \eqref{sys1}--\eqref{sys5}.

The singular points of the dynamical system $M_{0}$ are determined by a system of algebraic equations (see, for example, \cite{Bogoyav,Bautin}):
\begin{equation} \label{Eq__30_}
M_{0}:\quad F_{i} (x_{1} ,\ldots ,x_{n} )=0,\quad i=\overline{1,n}.
\end{equation}
From equations \eqref{sys1}, \eqref{sys3} we immediately obtain for singular points:
\begin{equation}\label{Zz0}
Z=0;\quad z=0.
\end{equation}
Then the equations $Z'=0,z'=0$ \eqref{sys2} -- \eqref{sys4} take the form:
\[
V'_\Phi +\gamma v'_\varphi=0; \quad -v'_\varphi+\gamma V'_\Phi=0.
\]
Since the determinant of this homogeneous system of algebraic equations linear with respect to derivative potentials is equal to $-(1+\gamma^2)\not=0$, the system of these equations has only a trivial solution:
\begin{eqnarray}
\label{Eq_Phi(0)}
V'_\Phi=0\Rightarrow\Phi(m^2-\alpha \Phi^{2})=0;\\
\label{Eq_varphi(0)}
v'_\varphi=0\Rightarrow\varphi(\mu ^{2} -\beta \varphi^{2})=0,
\end{eqnarray}
i.e., as we noted above, the singular points of a dynamic system are located at the extremum points of potential energy. Thus, the singular points of a cosmological asymmetric scalar doublet with kinetic coupling coincide with the singular points of the corresponding doublet with minimal coupling. This is an important property that greatly facilitates research. Note that the equation $F_5=0$ corresponding to \eqref{sys5} is identically satisfied by the solutions \eqref{Zz0}. As a result, since the kinetic interaction constant $\gamma$ does not affect the value of the effective energy $E$ at singular points, all singular points are accessible \cite{TMF20}.

So, the singular points of the dynamic system \eqref{sys1} -- \eqref{sys5} are determined by the system of algebraic equations:
\begin{eqnarray}\label{Point}
Z=0,& \Phi(m^2-\alpha\Phi^2)=0,\nonumber\\
z=0,&\varphi(\mu^2-\beta\varphi^2)=0,\\\nonumber
3H^2=&\displaystyle-\frac{\alpha\Phi^4}{4}+\frac{m^2\Phi^2}{2}-\frac{\beta\varphi^4}{4}+\frac{\mu^2\varphi^2}{2}+\Lambda.
\end{eqnarray}
Thus, there are 18 singular points in total.

\noindent 1. \textit{$M_{0,0}^{\pm} $}: For all values of $\alpha$ and $\beta$, the system of algebraic equations \eqref{Point} always has 2 solutions:
\begin{equation*}
\Phi=0;\; Z=0;\; \varphi=0;\; z=0;\; H=\pm\frac{\sqrt{3\Lambda}}{3}.
\end{equation*}
Thus, we have 2 symmetrical points located on the $OH$ axis:
\begin{equation} \label{Eq__32_}
M_{0,0}^{\pm} :\left(0,0,0,0,
\pm\frac{\sqrt{3\Lambda}}{3}
\right).
\end{equation}
Substituting the resulting solution \eqref{Eq__32_} into the condition \eqref{E>=0_2}, we obtain the necessary condition for the reality of solutions at a singular point:
\[\Lambda \ge 0.\]
\noindent 2. \textbf{$M_{0,\pm 1}^{\pm} $}: For all $\alpha$ and $\beta>0$ we have 4 more solutions that are pairwise symmetric with respect to $\varphi $ and $H$:
\begin{equation*}
\hspace{-0.5cm}\Phi=0;\; Z=0;\; \varphi =\pm \frac{\mu }{\sqrt{ \beta} };\; z=0;\; H=\pm\frac{\sqrt{3\Lambda_{\beta}}}{3}
\end{equation*}
\begin{equation}\label{Eq__34_}
\Rightarrow M_{0,\pm 1}^{\pm} :\left(0,0,\pm \frac{\mu}{\sqrt{\beta}},0,
\pm\frac{\sqrt{3\Lambda_{\beta}}}{3}\right),
\end{equation}
where
$$
\Lambda_{\beta}\equiv\Lambda+\frac{\mu^4}{4\beta}.
$$
A necessary condition for the reality of solutions at singular points \textbf{$M_{0,\pm 1}^{\pm} $}:
\begin{equation}\label{sigma1^2}
\Lambda_{\beta} \geqslant 0.
\end{equation}
\noindent 3.~\textbf{$M_{\pm 1,0}^{\pm} $}: For all $\beta$ and $\alpha>0$ there are 4 more points that are pairwise symmetric in $\Phi $ and $H$:
\begin{equation*}
\hspace{-0.5cm}\Phi =\pm \frac{m}{\sqrt{\alpha} } ;\; Z=0;\; \varphi=0;\; z=0;\; H=\pm\frac{\sqrt{3\Lambda_{\alpha}}}{3}
\end{equation*}
\begin{equation}\label{Eq__36_}
\Rightarrow M_{\pm 1,0}^{\pm}\left(\pm \frac{m}{\sqrt{\alpha} },0,0,0,\pm\frac{\sqrt{3\Lambda_{\alpha}}}{3}\right),
\end{equation}
where
$$
\Lambda_{\alpha}\equiv\Lambda+\frac{m^4}{4\alpha}.
$$
A necessary condition for the reality of solutions at singular points \textbf{$M_{\pm 1,0}^{\pm} $}:
\begin{equation}\label{sigma1^2}
\Lambda_{\alpha}  \geqslant 0.
\end{equation}
\noindent 4.~$M_{\pm 1,\pm 1}^{\pm}$: For $\alpha>0$ and $\beta>0$ there are 4 points that are pairwise symmetric in $\Phi $ and 4 points that are pairwise symmetric in $\varphi $:
\begin{equation*}
\hspace{-0.5cm}\Phi =\pm \frac{m}{\sqrt{\alpha} } ;\; Z=0;\; \varphi=\pm\frac{\mu}{\sqrt{\beta}};\; z=0;\; H=\pm\frac{\sqrt{3\Lambda_0}}{3}
\end{equation*}
\begin{equation}\label{Eq__38_}
\Rightarrow M_{\pm 1,\pm 1}^{\pm}\left(\pm \frac{m}{\sqrt{\alpha} },0,\pm\frac{\mu}{\sqrt{\beta}},0,\pm\frac{\sqrt{3\Lambda_0}}{3}\right).
\end{equation}
A necessary condition for the reality of solutions at singular points \textbf{$M_{\pm 1,\pm 1}^{\pm} $}:
\begin{equation}\label{sigma3^2}
\Lambda_0 \geqslant 0.
\end{equation}

Thus, the following statement is true:
\begin{stat}\label{stat3}
The dynamic system \eqref{sys1} -- \eqref{sys5} can have 2 ($M_{0,0}^{\pm}$), 6 (\{$M_{0,0}^{\pm}$ , $M_{0,\pm 1}^{\pm} $\} or \{$M_{0,0}^{\pm}$, $M_{\pm 1,0}^{\pm} $ \})
or 18 (\{$M_{0,0}^{\pm}, M_{0,\pm 1}^{\pm}, M_{\pm 1,0}^{\pm}$, $ M_{\pm 1,\pm 1}^{\pm}$\}) singular points whose coordinates coincide with the coordinates of the corresponding singular points of the dynamical system without interaction of components ($\gamma\equiv0$, see \cite{TMF20}).
\end{stat}

\subsection{The character of singular points}
The character of the singular points of a dynamical system, established by the values of the eigenvalues of the characteristic matrix at these points, determines the asymptotic behavior of the dynamical system near these points. The equations that determine the eigenvectors $\mathbf{u}_i$ and eigenvalues $\lambda_i$ of the matrix of the dynamical system are
\begin{eqnarray}\label{eigen_vectors}
\bigl(\mathbf{A}_M-\lambda_i\mathbf{E}\bigr)\mathbf{u}_i=0;\\
\label{eigen_values}
\mathrm{Det}\bigl(\mathbf{A}_M-\lambda_i\mathbf{E}\bigr)=0,
\end{eqnarray}
where $\mathbf{E}$ is the identity matrix.
According to the qualitative theory of differential equations (see, for example, \cite{Bogoyav,Bautin} the radius vector of the phase trajectory $\mathrm{r}(t)=(x_1(t),\ldots,x_n(t))$ in the neighborhood of the singular point $M^\alpha(x_1^{(\alpha)},\ldots,x^{(\alpha)}_n)$ is described by the equation:
\begin{equation}\label{r(t)}
\mathbf{r}(t)\simeq \mathbf{r}^{(\alpha)}+\Re\biggl(\sum\limits_{j=1}^n C_j \mathbf{u}^{(\alpha)}_j e^{i\lambda^{(\alpha)}_jt}\biggr),
\end{equation}
where $C_j$ are arbitrary constants determined by the initial conditions, $\lambda^{(\alpha)}_j$ are the eigenvalues of the dynamic system matrix $\mathbf{A}(M_\alpha)$, $\mathbf{u }^{(\alpha)}_j$ are the eigenvectors of this matrix corresponding to the eigenvalues of $\lambda^{(\alpha)}_j$. The estimation formula (\ref{r(t)}) is a fairly good approximation of the phase trajectory.

The characteristic matrix of the dynamic system \eqref{sys1} -- \eqref{sys5} has the form
{\small
$$\hspace{-4.5cm}\mathbf{A}_M\equiv\left\| \frac{\partial F_{i} }{\partial x_{k} } \right\|_M= $$
\vspace{-4mm}
\begin{eqnarray}\label{matrica0}
\left(\begin{array}{ccccc}
 {0} & {1} & {0} & {0} & {0}\\
 {\displaystyle\frac{\partial F_2}{\partial \Phi} } &-3H & {\displaystyle-\gamma\frac{\partial F_4}{\partial \varphi} } & {0} &  -3Z\\
 {0} & {0} & {0} & {1}&0 \\
 {\displaystyle\gamma\frac{\partial F_2}{\partial \Phi} } & {0} & {\displaystyle\frac{\partial F_4}{\partial \varphi} } & -3H&-3z\\
 0&{\displaystyle\frac{\partial F_5}{\partial Z} }&0&{\displaystyle\frac{\partial F_5}{\partial z} }&-6H
 \end{array}\right)_M\!,
\end{eqnarray}}
where the partial derivatives are equal
\begin{equation}
\hspace{-0.5cm}\begin{array}{ll}
{\displaystyle\frac{\partial F_2}{\partial \Phi} =-\frac{m^2-3\alpha\Phi^2}{1+\gamma^2}};&{\displaystyle\frac{\partial F_4}{\partial \varphi}=\frac{\mu^2-3\beta\varphi^2 }{1+\gamma^2}};\vspace{0.3cm}\\
{\displaystyle\frac{\partial F_5}{\partial Z} =-Z-\gamma z};&{\displaystyle\frac{\partial F_5}{\partial z}=z-\gamma Z}.
\end{array}
\end{equation}
Taking into account the definition of singular points $Z=z=0$, the matrix \eqref{matrica0} takes the form:
{\small
\begin{equation}\label{matrica}
\hspace{-0.7cm}\mathbf{A}_{M_0} =\left(\begin{array}{ccccc}
 {0} & {1} & {0} & {0} & {0}\\
 {\displaystyle\frac{\partial F_2}{\partial \Phi} } &-3H & {\displaystyle-\gamma\frac{\partial F_4}{\partial \varphi} } & {0} &  0\\
 {0} & {0} & {0} & {1}&0 \\
 {\displaystyle\gamma\frac{\partial F_2}{\partial \Phi} } & {0} & {\displaystyle\frac{\partial F_4}{\partial \varphi} } & -3H&0\\
 0&0&0&0&-6H
 \end{array}\right).
\end{equation}
}
The characteristic equation for the matrix $\mathbf{A}_M$ has the form:
{\small\begin{equation}\label{charac}
\hspace{-6mm}\begin{array}{l}
\left(-6H-\lambda\right)\left(\lambda^2(3H+\lambda)^2-\right.\\[8pt]
\left.\left(\dfrac{\partial F_2}{\partial \Phi}+\dfrac{\partial F_4}{\partial \varphi}\right)\lambda(3H+\lambda)+(1+\gamma^2)\dfrac{\partial F_2}{\partial \Phi}\dfrac{\partial F_4}{\partial \varphi}\right)=0,
\end{array}
\end{equation}}
which gives the following eigenvalues:
{\small\begin{equation}\label{lambdas}
\hspace{-6mm}\begin{array}{l}
\lambda_{(5)}=-6H,\\
\lambda_{(1,2,3,4)}=-\dfrac{3H}{2}\pm\left[\left(\dfrac{3H}{2}\right)^2+\dfrac12\left(\dfrac{\partial F_2}{\partial \Phi}+\dfrac{\partial F_4}{\partial \varphi}\right)\right.\\
\left. \pm\dfrac12\sqrt{\left(\dfrac{\partial F_2}{\partial \Phi}+\dfrac{\partial F_4}{\partial \varphi}\right)^2-4\dfrac{\partial F_2}{\partial \Phi}\dfrac{\partial F_4}{\partial \varphi}(1+\gamma^2)}\right]^{\frac12}.
\end{array}
\end{equation}}
Before writing out the values of the eigenvalues at singular points, we make the following remark. The coordinates of the eigenvectors at all singular points in the limit $\gamma\to0$, with correct numerical calculations, transform into the corresponding eigenvectors of the model without interaction of the components ($\gamma\equiv0$)\footnote{To do this, you must first substitute the numerical values of the fundamental parameters $\mathbf{P}$ \eqref{P} into the characteristic matrix $\mathbf{A}$ \eqref{matrica} and only then calculate the eigenvectors.}. In this regard, we will limit ourselves to giving expressions for the components of eigenvectors in the space ${\mathbb R}^{5}$ \eqref{R5} for the case $\gamma=0$ for arbitrary eigenvalues $\lambda_i$ \cite{TMF20 }:
\begin{equation}\label{u}
\begin{array}{l}
\textbf{u}_{(1,2)}=[1,\lambda_{(1,2)},0,0,0],\\[3pt]
\textbf{u}_{(3,4)}=[0,0,1,\lambda_{(3,4)},0],\\[3pt]
\mathbf{u}_{(5)}=[0,0,0,0,1].
\end{array}, \qquad (\gamma=0).
\end{equation}
Thus, for $\gamma\to0$ the vectors $\textbf{u}_{(1,2)}$ are in the two-dimensional subspace $\Pi_{\Phi}$, the vectors $\textbf{u}_{(3 ,4)}$ -- in the subspace $\Pi_{\varphi}$, and the vector $\textbf{u}_{(5)}$ in the one-dimensional subspace $\mathbb{R}^1=\{H\} $.

Below we briefly list the results of a qualitative analysis of the dynamic system \eqref{sys1} -- \eqref{sys5}. Calculations show that all singular points of a dynamical system are divided into 4 groups, the character of the points within each group is the same (for details, see \cite{TMF20}).

\subsubsection*{Singular points $M_{0,0}^{\pm} $:}
The first group includes 2 singular points:
\[M_{0,0}^{\pm}\left(0,0,0,0,\pm\frac{\sqrt{3\Lambda}}{3}\right); \quad \alpha\in \mathbb{R},\quad \beta\in \mathbb{R}\]
The system matrix \eqref{matrica} at zero singular points \eqref{Eq__32_} for all $\alpha$ and $\beta$ will take the form:
{\small
\[
\mathbf{A}_{0} \equiv \mathbf{A}(M_{0,0}^{\pm} )=\]
\vspace{-4mm}
\!\!\!\!\!\!\!\!\!\!\!\!\!\!\!\!\!\!\[\left(\begin{array}{ccccc}
 {0} & {1} & {0} & {0} & {0}\\
 {-\frac{m^2}{1+\gamma^2} } &\mp\sqrt{3\Lambda} & {-\gamma\frac{\mu^2}{1+\gamma^2}} & {0} &  0\\
 {0} & {0} & {0} & {1}&0 \\
 {-\gamma\frac{m^2}{1+\gamma^2} } & {0} & {\frac{\mu^2}{1+\gamma^2} } & \mp\sqrt{3\Lambda}&0\\
 0&0&0&0&\mp2\sqrt{3\Lambda}
 \end{array}\right),
\]}
its determinant is equal\footnote{The sequences of signs in the expressions for the determinant and the characteristic matrix are the same.}:
\[
\Delta (\mathbf{A}_{0} )=\pm\frac{2\sqrt{3\Lambda}m^2 \mu ^{2}}{1+\gamma^2} .
\]
In the case of an expanding Universe, which corresponds to the choice of a positive sign $H$ in the formula \eqref{Eq__32_}, the characteristic equation of the system at the singular point $M_{0,0}^{+} $ takes the form
{\small
\[\begin{array}{l}
\!\!\!\!\!\!\!\left(-2\sqrt{3\Lambda}-\lambda\right)\times\\\!\!\!\!\!\!\!\left(\lambda^2(\sqrt{3\Lambda}+\lambda)^2+\dfrac{m^2-\mu^2}{1+\gamma^2}\lambda(\sqrt{3\Lambda}+\lambda)-\dfrac{m^2\mu^2}{1+\gamma^2}\right)=0,
\end{array}\]}
from where we get the eigenvalues:
{\small
\begin{equation} \label{lambda_0}
\begin{array}{l}
\hspace{-5mm}\lambda_{0\,(5)}=-2\sqrt{3\Lambda},\\
\hspace{-5mm}
\lambda_{0\,(1,2)}=-\sqrt{\dfrac{3\Lambda}{4}}\pm\\
\hspace{-5mm}\sqrt{\displaystyle\dfrac{3\Lambda}{4}-\dfrac{m^2-\mu^2+\sqrt{\left(m^2+\mu^2\right)^2+4\gamma^2m^2\mu^2}}{2(1+\gamma^2)}},\\
\hspace{-5mm}\lambda_{0\,(3,4)}=-\sqrt{\dfrac{3\Lambda}{4}}\pm\\
\hspace{-5mm}\sqrt{\displaystyle\dfrac{3\Lambda}{4}-\dfrac{m^2-\mu^2-\sqrt{\left(m^2+\mu^2\right)^2+4\gamma^2m^2\mu^2}}{2(1+\gamma^2)}}.
\end{array}
\end{equation}}

\subsubsection*{Singular points $M_{0,\pm 1}^{\pm} $: }
The second group, under the condition $\beta>0$ and $\forall\alpha$, includes 4 pairwise symmetric singular points (\ref{Eq__34_}):
\[M_{0,\pm 1}^{\pm} :\left(0,0,\pm \frac{\mu}{\sqrt{\beta}},0,
\pm\frac{\sqrt{3\Lambda_{\beta}}}{3}\right),\quad
\Lambda_{\beta}\equiv\Lambda+\frac{\mu^4}{4\beta}.
\]
System matrix \eqref{matrica} at singular points $M_{0,\pm 1}^{\pm} $:
{\small
\[
\hspace{-2mm}\mathbf{A}_{01} \equiv \mathbf{A}(M_{0,\pm1}^{\pm} )=\]
\vspace{-4mm}
\[\hspace{-3mm}\left(\begin{array}{ccccc}
 {0} & {1} & {0} & {0} & {0}\\
 {-\frac{m^2}{1+\gamma^2} } &\mp\sqrt{3\Lambda_{\beta}} & {\gamma\frac{2\mu^2}{1+\gamma^2}} & {0} &  0\\
 {0} & {0} & {0} & {1}&0 \\
 {-\gamma\frac{m^2}{1+\gamma^2} } & {0} & {-\frac{2\mu^2}{1+\gamma^2} } & \mp\sqrt{3\Lambda_{\beta}}&0\\
 0&0&0&0&\mp2\sqrt{3\Lambda_{\beta}}
 \end{array}\right),
\]}

\hspace{-7mm}its determinant is equal to:
\[
\Delta (\mathbf{A}_{01} )=\mp\frac{4\sqrt{3\Lambda_{\beta}}m^2 \mu ^{2}}{1+\gamma^2} .
\]
The characteristic equation of the system at singular points $M_{0,\pm 1}^{+} $ has the form:
{\small
\[\hspace{-8mm}\begin{array}{l}
\left(-2\sqrt{3\Lambda_{\beta}}-\lambda\right)\times\\\left(\lambda^2(\sqrt{3\Lambda_{\beta}}+\lambda)^2+\dfrac{m^2+2\mu^2}{1+\gamma^2}\lambda(\sqrt{3\Lambda_{\beta}}+\lambda)+\dfrac{2m^2\mu^2}{1+\gamma^2}\right)=0,
\end{array}\]}
\hspace{-2mm}from where we get the eigenvalues:
{\small
\begin{equation} \label{lambda_01}
\hspace{-7mm}\begin{array}{l}
\lambda_{01\,(5)}=-2\sqrt{3\Lambda_{\beta}},\\
\lambda_{01\,(1,2)}=-\sqrt{\dfrac{3\Lambda_{\beta}}{4}}\pm\\
\sqrt{\dfrac{3\Lambda_{\beta}}{4}-\dfrac{m^2+2\mu^2+\sqrt{\left(m^2-2\mu^2\right)^2-8\gamma^2m^2\mu^2}}{2(1+\gamma^2)}},\\
\lambda_{01\,(3,4)}=-\sqrt{\dfrac{3\Lambda_{\beta}}{4}}\pm\\
\sqrt{\dfrac{3\Lambda_{\beta}}{4}-\dfrac{m^2+2\mu^2-\sqrt{\left(m^2-2\mu^2\right)^2-8\gamma^2m^2\mu^2}}{2(1+\gamma^2)}}.
\end{array}
\end{equation}}
\subsubsection*{Singular points $M_{\pm 1,0}^{\pm} $: }
The third group, under the condition $\alpha>0$ and $\forall\beta$, includes 4 pairwise symmetric singular points \eqref{Eq__36_}:
\[M_{\pm 1,0}^{\pm} :\left(\pm \frac{m}{\sqrt{\alpha}},0,0,0,
\pm\frac{\sqrt{3\Lambda_{\alpha}}}{3}\right),\;
\Lambda_{\alpha}\equiv\Lambda+\frac{m^4}{4\alpha}.
\]
System matrix \eqref{matrica} at singular points $M_{\pm 1.0}^{\pm} $:
{\small
\[\hspace{-3mm}
\mathbf{A}_{10} \equiv \mathbf{A}(M_{\pm1,0}^{\pm} )=\]
\vspace{-6mm}
\[\hspace{-5mm}\left(\begin{array}{ccccc}
 {0} & {1} & {0} & {0} & {0}\\
 {-\frac{4m^2}{1+\gamma^2} } &\mp\sqrt{3\Lambda_{\alpha}} & {-\gamma\frac{\mu^2}{1+\gamma^2}} & {0} &  0\\
 {0} & {0} & {0} & {1}&0 \\
 {-\gamma\frac{4m^2}{1+\gamma^2} } & {0} & {\frac{\mu^2}{1+\gamma^2} } & \mp\sqrt{3\Lambda_{\alpha}}&0\\
 0&0&0&0&\mp2\sqrt{3\Lambda_{\alpha}}
 \end{array}\right),
\]}
its determinant is equal to:
\[
\Delta (\mathbf{A}_{10} )=\pm\frac{8\sqrt{3\Lambda_{\alpha}}m^2 \mu ^{2}}{1+\gamma^2} .
\]
The characteristic equation of the system at singular points $M_{\pm 1,0}^{+} $ has the form:
{\small
\[\hspace{-4mm}\begin{array}{l}
\!\!\!\!\!\!\!\left(-2\sqrt{3\Lambda_{\alpha}}-\lambda\right)\times\\\!\!\!\!\!\!\!\left(\lambda^2(\sqrt{3\Lambda_{\alpha}}+\lambda)^2+\dfrac{4m^2-\mu^2}{1+\gamma^2}\lambda(\sqrt{3\Lambda_{\alpha}}+\lambda)-\dfrac{4m^2\mu^2}{1+\gamma^2}\right)=0,
\end{array}\]}
\hspace{-2mm}from where we get the eigenvalues:
{\small
\begin{equation} \label{lambda_10}\hspace{-7mm}
\begin{array}{l}
\lambda_{10\,(5)}=-2\sqrt{3\Lambda_{\alpha}},\\
\lambda_{10\,(1,2)}=-\sqrt{\dfrac{3\Lambda_{\alpha}}{4}}\pm\\
\sqrt{\dfrac{3\Lambda_{\alpha}}{4}+
\dfrac{\mu^2+2m^2-\sqrt{\left(\mu^2-2m^2\right)^2-8\gamma^2m^2\mu^2}}{2(1+\gamma^2)}},\\
\lambda_{10\,(3,4)}=-\sqrt{\dfrac{3\Lambda_{\alpha}}{4}}\pm\\
\sqrt{\dfrac{3\Lambda_{\alpha}}{4}+
\dfrac{\mu^2+2m^2+\sqrt{\left(\mu^2-2m^2\right)^2-8\gamma^2m^2\mu^2}}{2(1+\gamma^2)}}.
\end{array}
\end{equation}}
\subsubsection*{Singular points $M_{\pm 1,\pm 1}^{\pm} $: }
The fourth group, under the condition $\alpha>0$ and $\beta>0$, includes another 8 pairwise symmetric singular points \eqref{Eq__38_}\footnote{We used the relation \eqref{L->L0}} here:
\[M_{\pm 1,\pm 1}^{\pm} :\left(\pm \frac{m}{\sqrt{\alpha}},0,\pm \frac{\mu}{\sqrt{\beta}},0,
\pm\frac{\sqrt{3\Lambda_{\alpha\beta}}}{3}\right),\]
\[\Lambda_{\alpha\beta}\equiv\Lambda+\frac{m^4}{4\alpha}+\frac{\mu^4}{4\beta}\equiv \Lambda_0.
\]
System matrix \eqref{matrica} at singular points $M_{\pm 1,\pm 1}^{\pm} $:
{\small
\[\hspace{-3mm}
\mathbf{A}_{11} \equiv \mathbf{A}(M_{\pm1,\pm1}^{\pm} )=\]
\vspace{-6mm}
\[\hspace{-5mm}\left(\begin{array}{ccccc}
 {0} & {1} & {0} & {0} & {0}\\
 {-\frac{4m^2}{1+\gamma^2} } &\mp\sqrt{3\Lambda_{0}} & {\gamma\frac{2\mu^2}{1+\gamma^2}} & {0} &  0\\
 {0} & {0} & {0} & {1}&0 \\
 {-\gamma\frac{4m^2}{1+\gamma^2} } & {0} & {-\frac{2\mu^2}{1+\gamma^2} } & \mp\sqrt{3\Lambda_{0}}&0\\
 0&0&0&0&\mp2\sqrt{3\Lambda_{0}}
 \end{array}\right),
\]}

\hspace{-7mm}its determinant is equal to:
\[
\Delta (\mathbf{A}_{11} )=\mp\frac{16\sqrt{3\Lambda_{\alpha\beta}}m^2 \mu ^{2}}{1+\gamma^2} .
\]

The characteristic equation of the system at singular points $M_{\pm 1,\pm1}^{+} $ has the form:
{\small
\[\hspace{-8mm}\begin{array}{l}
\left(-2\sqrt{3\Lambda_{\alpha\beta}}-\lambda\right)\times\\\left(\lambda^2(\sqrt{3\Lambda_{\alpha\beta}}+\lambda)^2+\frac{4m^2+2\mu^2}{1+\gamma^2}\lambda(\sqrt{3\Lambda_{\alpha\beta}}+\lambda)+\frac{8m^2\mu^2}{1+\gamma^2}\right)=0,
\end{array}\]}
\hspace{-2mm}from where we get the eigenvalues:
{\small
\begin{equation} \label{lambda_11}
\hspace{-6mm}
\begin{array}{l}
\lambda_{11\,(5)}=-2\sqrt{3\Lambda_{\alpha\beta}},\\
\lambda_{11\,(1,2)}=-\sqrt{\dfrac{3\Lambda_{\alpha\beta}}{4}}\pm\\
\sqrt{\dfrac{3\Lambda_{\alpha\beta}}{4}+\dfrac{m^2-\mu^2+\sqrt{\left(m^2+\mu^2\right)^2+4\gamma^2m^2\mu^2}}{1+\gamma^2}},\\
\lambda_{11\,(3,4)}=-\sqrt{\dfrac{3\Lambda_{\alpha\beta}}{4}}\pm\\
\sqrt{\dfrac{3\Lambda_{\alpha\beta}}{4}+\dfrac{m^2-\mu^2-\sqrt{\left(m^2+\mu^2\right)^2+4\gamma^2m^2\mu^2}}{1+\gamma^2}}.
\end{array}
\end{equation}}
It can be verified that, in the limit
\begin{equation}\label{gamma->0}
\gamma\to0
\end{equation}
expressions for the eigenvalues \eqref{lambda_0}, \eqref{lambda_01}, \eqref{lambda_10} and \eqref{lambda_11} of the characteristic matrix $\mathbf{A}$ \eqref{matrica} at all singular points of the dynamical system $M_{ 0,0}^{\pm}$, $M_{0,\pm 1}^{\pm}$, $M_{\pm 1,0}^{\pm}$ and $M_{\pm 1, \pm 1}^{\pm} $ go into the corresponding expressions for the eigenvalues of the dynamical system of an asymmetric scalar doublet with non-interacting components \cite{TMF20}. Therefore, the character of these singular points is preserved in the passage to the limit.

Here and in what follows, the following notations are used: $\mathbf{A}$ -- attraction, $\mathbf{R}$ -- repulsion, $\mathbf{S}$ -- saddle. Here and in what follows, we somewhat simplify the characteristics of singular points, preserving only their properties of attraction and repulsion, since it is these properties that significantly influence the qualitative analysis of a dynamic system. We will denote the total characteristics of singular points by an ordered pair of symbols, for example, [\textbf{A, S}].
\begin{center}
\refstepcounter{table} {\large Tabl. \thetable. Character (type) of singular points in subspaces $[\Sigma_\Phi,\Sigma_\varphi]$}\\[12pt]
\begin{tabular}{|c|c|}
\hline
Singular points & Type of points \\
\hline
$M_{0,0}^{+}$ &  $[\mathbf{A},\ \mathbf{S}]$ \\
\hline
$M_{0,0}^{-}$ &  $[\mathbf{R},\ \mathbf{S}]$\\
\hline
$M_{0,\pm1}^{+}$ &  $[\mathbf{A},\ \mathbf{A}]$ \\
\hline
$M_{0,\pm1}^{-}$ &  $[\mathbf{R},\ \mathbf{R}]$ \\
\hline
$M_{\pm1,0}^{+}$ & $[\mathbf{S},\ \mathbf{S}]$ \\
\hline
$M_{\pm1,0}^{-}$ &  $[\mathbf{S},\ \mathbf{S}]$ \\
\hline
$M_{\pm1,\pm1}^{+}$ & $[\mathbf{S},\ \mathbf{A}]$ \\
\hline
$M_{\pm1,\pm1}^{-}$ & $[\mathbf{S},\ \mathbf{R}]$ \\
\hline
\end{tabular}
\end{center}
\subsubsection*{Comparison with a model without interaction between components}

By directly calculating expressions for the eigenvalues \eqref{lambdas} at the specified singular points $M_{0,0}^{\pm} $, $M_{0,\pm 1}^{\pm} $, $M_{\pm 1,0}^{\pm} $ and $M_{\pm 1,\pm 1}^{\pm}$ it can be shown that the signs of the real parts of the eigenvalues coincide with the signs of the corresponding eigenvalues in the model without interaction of components (see, \cite{TMF20}). Thus, the following statement is true:
\begin{stat}\label{stat3}
The character of the singular points of the dynamical system \eqref{sys1} -- \eqref{sys5} coincides with the character of the corresponding singular points of the dynamical system without interaction of components ($\gamma\equiv0$, see \cite{TMF20}).
\end{stat}

\section{Asymptotic behavior\newline near singularities}
By analogy with the work of \cite{Ignat_Sasha_Dima2}, we study the asymptotic behavior of the cosmological model in the vicinity of singular points $t_s$, which are determined by the scale factor being equal to zero
\begin{equation}\label{a=0}
a(t_s)=0\Rightarrow \xi(t_s)=-\infty.
\end{equation}
Since, however, in contrast to the indicated work, in this work the right-hand sides of the dynamical system \eqref{sys1} - \eqref{sys5} (or \eqref{sys1a} - \eqref{sys5a}) as well as the left-hand side of the hypersurface equation \eqref{Hip_Ein} do not depend explicitly on the scale function $\xi(t)$, we will define these singular points in a different way. As the results of numerical modeling show at singular points of the $t_s$ model, if they exist, the relations are satisfied\footnote{These results will be presented in the second part of the article.}
\begin{eqnarray}\label{sing_points_t}
|H(t_s)|\to-\infty,& |Z(t_s)|\to \infty,|z(t_s)|\to \infty,\nonumber\\
|\Phi(t_s)|\to \infty, & |\varphi(t_s)|\to \infty. \hspace{1cm}
\end{eqnarray}
Thus, at singular points $t_s$ all dynamic functions of the model are also singular. Then the Einstein hypersurface equation \eqref{Hip_Ein} implies estimates near the singularity:
\begin{eqnarray}\label{H=Z=z}
|H(t_s)|\sim |Z(t_s)|\sim z(t_s)|,\nonumber\\
|\Phi(t_s)|\sim |\varphi(t_s)|\sim |H^{1/2}(t_s)|.
\end{eqnarray}
Using the estimates \eqref{H=Z=z} in the dynamical system \eqref{sys0a} -- \eqref{sys5a}, we reduce it near the singularity $t\to t_s$ to the form
\begin{eqnarray}
\label{sys0a0}
\dot{\xi}= H ;& \displaystyle \Rightarrow \frac{d}{dt}=H\frac{d}{d\xi}; \\
\label{sys1a0}
\dot{\Phi}=  Z; & \displaystyle \dot{\varphi}=z;\\
\label{sys2a0}
\dot{Z}  =  -3HZ; & \displaystyle \dot{z} = -3Hz;  \\
\label{sys3a0}
\dot{H} = & \displaystyle -\frac{1}{2}Z^2+\frac{1}{2}z^2-\gamma Zz.
\end{eqnarray}
Using the substitution \eqref{sys0a0} in the equations \eqref{sys2a0}, we find solutions to these equations near the singularity:
\begin{eqnarray}\label{Z=,z=}
Z\backsimeq C_1 \mathrm{e}^{-3\xi}; & \displaystyle z\backsimeq C_2 \mathrm{}e^{-3\xi}.
\end{eqnarray}
Using the substitution \eqref{sys0a0} and the found solutions \eqref{Z=,z=} in the equation \eqref{sys3a0}, we obtain its solution near the singularity:
\begin{eqnarray}\label{H2}
H^2=\frac{1}{6}\bigl(C_1^2-C_2^2+2\gamma C_1C_2\bigr)\mathrm{e}^{-6\xi}.
\end{eqnarray}

Note that from \eqref{H2}, in particular, follows the necessary condition for the existence of a singularity
\begin{eqnarray}\label{H2>0}
H^2\geqslant0 \Rightarrow C_1^2-C_2^2+2\gamma C_1C_2\geqslant0.\nonumber
\end{eqnarray}
Substituting $H(\xi)$ from \eqref{H2} into \eqref{sys0a0} and integrating this equation, we find the asymptotic value of the scale factor $a(t)=\exp(\xi(t))$ near the singularity:
\begin{equation}
a(t)=\left[\frac{3}{2}\bigl(C_1^2-C_2^2+2\gamma C_1C_2\bigr)\right]^{\frac{1}{3}}|t-t_s|^{\frac{1}{3}}
\end{equation}
Then using the expression for the derivative from \eqref{sys0a0} into \eqref{sys1a0} and substituting it into the resulting $H(\xi)$ from \eqref{H2}, we find the asymptotic value of the potentials $\Phi(t)$ and $\varphi(t)$ near the singularity:
\begin{eqnarray}
\Phi=\pm\dfrac{C_1}{\sqrt{C_1^2-C_2^2+2\gamma C_1C_2}}\xi;\\
\varphi=\pm\dfrac{C_2}{\sqrt{C_1^2-C_2^2+2\gamma C_1C_2}}\xi.
\end{eqnarray}

Finally, using the relations \eqref{sys3a0}, \eqref{Z=,z=} and \eqref{H2}, we find the asymptotic value of the cosmological acceleration \eqref{Omega} near the singularity
\begin{equation}\label{varkappa=1}
\Omega(t_s)\backsimeq -2\Rightarrow \kappa(t_s)\backsimeq 1,
\end{equation}
i.e., near cosmological singularities, the effective equation of state of matter is extremely rigid. Note that the value of cosmological acceleration near singularities in the cosmological model with binary scalar charged fermions is equal to $\Omega=-1$, which corresponds to the ultra-relativistic state of matter near the singularity $\kappa=1/3$ \cite{Ignat_Sasha_Dima2}.

Note also that we could obtain the last result \eqref{varkappa=1} more simply, using the estimate \eqref{H=Z=z} near singularities in the expressions for the energy densities and pressures of individual components \eqref{E_cf-p_cf}. Thus, we find near the singularity
\begin{eqnarray}\label{e-p,t=t_s}
\varepsilon_c\backsimeq p_c\backsimeq \frac{Z^2}{16\pi}; & (t\to t_s)\\
\varepsilon_f\backsimeq p_f\backsimeq -\frac{z^2}{16\pi}; & \displaystyle \varepsilon_{cf}= p_{cf}= \frac{\gamma Zz}{8\pi},
\end{eqnarray}
from which it follows for all components of a dynamic system:
\begin{eqnarray}\label{O,kappa,t=t_s}
\kappa_c\backsimeq \kappa_f \backsimeq \kappa_{cf}=1, & (t\to t_s).
\end{eqnarray}
\section{Numerical modeling example}

So, a particular cosmological model $\mathbf{M}$ is determined, firstly, by the system of dynamic equations \eqref{sys0a}--\eqref{sys5a}, and secondly, by fundamental parameters \eqref{P} and thirdly, by the initial conditions \eqref{I}.%

Before moving on to the results of numerical simulation, we will demonstrate the numerical implementation of the property \ref{stat1} of invariance of a dynamic system with respect to a change in the sign of the constant $\gamma$.
\Fig{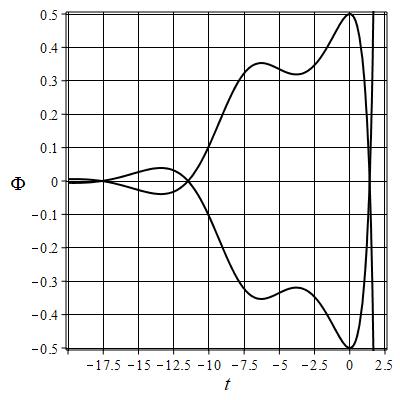}{8}{\label{Ign1} Evolution of $\Phi(t)$: $\mathbf{P_{+}}$, $\mathbf{I^{(1)}_{+}}$ and $\mathbf{P_{-}}$, $\mathbf{I^{(1)}_{-}}$.}
\Fig{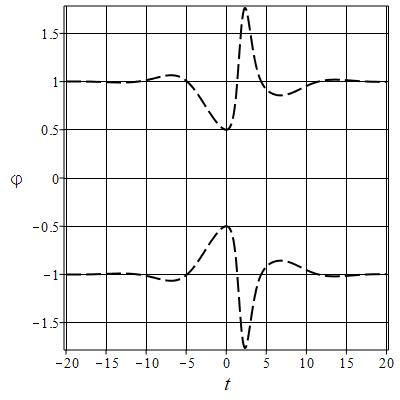}{8}{\label{Ign2} Evolution of  $\varphi(t)$: $\mathbf{P_{+}}$, $\mathbf{I^{(2)}_{+}}$ and $\mathbf{P_{-}}$, $\mathbf{I^{(2)}_{-}}$.}
In Fig.\ref{Ign1} -- \ref{Ign2} this property is demonstrated for an example model with parameters
\begin{equation}\label{P_{+}}
\mathbf{P_{\pm}}=[[1, 1,1,1],\pm 1,1]
\end{equation}
and initial conditions
\begin{eqnarray}\label{I_{+}}\nonumber
\!\!\!\!\!\mathbf{I^{(1)}_{\pm}}=[\pm 0.5,0,0,0,-1];
\mathbf{I^{(2)}_{\pm}}=[0,0,\pm 0.5,0,-1].\nonumber
\end{eqnarray}

Note that in this article we do not set the task of numerically studying models with real physical parameters $\mathbf{P}$, but only pursue the goal of demonstrating the general properties of such models. To apply the results of numerical simulation to a real physical model, you can use the  Property \ref{stat2} and use it to transform the fundamental constants and corresponding solutions according to the similarity laws \eqref{sim1} -- \eqref{sim3}.

\subsection{Singular points}
The coordinates of singular points and their characters in the subspaces $\Sigma_{\Phi}$ and $\Sigma_{\varphi}$ for the model with parameters $\mathbf{P_{+}}$ are indicated in Tabl.\ref{Tab2}, and their images are shown in Fig.\ref{Ign3} -- \ref{Ign4}.
\begin{center}\refstepcounter{table} \label{Tab2}
Tabl. \thetable.  Character of singular points \linebreak in the subspaces [$\Sigma_{\Phi},\Sigma_{\varphi}]$\\[6pt]
\begin{tabular}{|p{2cm}|p{3.3cm}|p{1.5cm}|}
\hline
Singular points & Coordinates & Type\\
\hline
$M_{0,0}^{+}$ & $[0,0,0,0.577]$ & $[\mathbf{A},\ \mathbf{S}]$ \\
\hline
$M_{0,0}^{-}$ & $[0,0,0,0,-0.577]$ &  $[\mathbf{R},\ \mathbf{S}]$ \\
\hline
$M_{\pm1,0}^{+}$ & $[\pm 1,0,0,0,0.645]$ &  $[\mathbf{S},\ \mathbf{S}]$\\
\hline
$M_{\pm1,0}^{-}$ & $[\pm 1,0, 0,0,-0.645]$ &  $[\mathbf{S},\ \mathbf{S}]$\\
\hline
$M_{0,\pm1}^{+}$ & $[0,0,\pm 1,0,0.645]$ &  $[\mathbf{A},\ \mathbf{A}]$ \\
\hline
$M_{0,\pm1}^{-}$ & $[0,0,\pm 1,0,-0.645]$ &  $[\mathbf{R},\ \mathbf{R}]$\\
\hline
$M_{\pm1,\pm1}^{+}$ & $[\pm 1,0,\pm 1,0, 0.707]$ &  $[\mathbf{S},\ \mathbf{A}]$\\
\hline
$M_{\pm1,\pm1}^{-}$ & $[\pm 1,0,\pm 1,0, -0.707]$  & $[\mathbf{S},\ \mathbf{R}]$\\
\hline
\end{tabular}
\end{center}
As can be seen in Fig.\ref{Ign3} -- \ref{Ign4}, singular points in both subspaces are located in close vertical pairs with the same character. This factor can lead to a wide variety of types of behavior of phase trajectories close to each other in the region of these special points.
\FigH{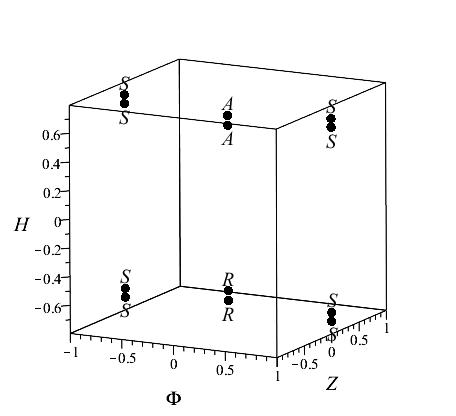}{7.5}{6.5}{\label{Ign3}
Singular points in the subspace $\Sigma_{\Phi}$ of the model with parameters $\mathbf{P_+}$ \eqref{P_{+}}.}

This situation is demonstrated in Fig.\ref{Ign4}, where the trajectories in the $\Sigma_\Phi$ subspace originate from one repulsive point in the lower part of the figure, then their graphs diverge, but later meet again at one attracting point in the upper part of the graph.
\Fig{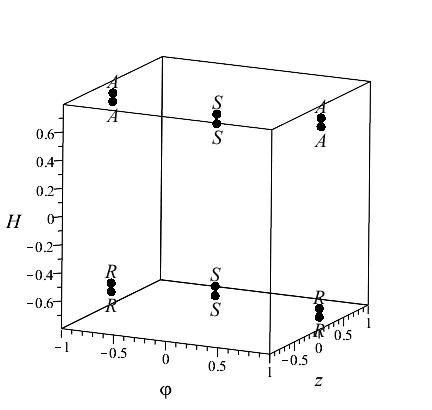}{8}{\label{Ign4}
Singular points in the subspace $\Sigma_{\varphi}$ of the model with parameters $\mathbf{P_+}$ \eqref{P_{+}}.}
\subsection{Evolution of scale functions and hypersurface sections Einstein-Higgs}
According to Tabl.\ref{Tab2} there is a single singular point, $M_{0,\pm1}^{+}$ (see \eqref{Eq__34_}), which is simultaneously attractive in the subspaces $\Sigma_{ \Phi}$ and $\Sigma_{\varphi}$.
In Fig. \ref{Ign6}--\ref{Ign8} shows graphs of the evolution of the scale functions $\xi(t)$ and $H(t)$ for the model with parameters $\mathbf{P_{+}}$ \eqref{P_{+}} and $\mathbf{P_{0}}=[[1, 1,1,1],0,1]$ and initial conditions close to an attracting singular point of type $[\mathbf{A},\mathbf{A}]$ $M_{0 ,-1}^{+}=$ $(0,0,-1,0,0.645)$:
\begin{eqnarray}\label{I_0}
\mathbf{I_0}=[0,0,-1,0,1]\Rightarrow (M_{0,-1}^{+}), \\
\label{I_1,2}
\!\!\!\mathbf{I_1}=[0,0,-0.99,0,1], \mathbf{I_2}=[0,0,-1.01,0,1].
\end{eqnarray}

This model has an infinite inflationary past ($H_{-\infty}\approx -0.645$) and an infinite inflationary future ($H_{+\infty}\approx 0.645$). The initial state $t=-\infty$ corresponds to unstable points: $M_{\pm1,0}^{-}$ -- type $[\mathbf{S},\mathbf{S}]$, $M_{0,\pm1}^{-}$ -- type $[\mathbf{R},\mathbf{R}]$, and the final state $t=+\infty$ - stable points of type $[\mathbf{A},\mathbf{A}]$ $M_{0,-1}^{+}$ (see Tabl.\ref{Tab2}).

Fig.\ref{Ign6} -- \ref{Ign8} shows the evolution of the scale function $\xi(t)=\ln(a(t))$ and the Hubble parameter $H(t)$ \eqref{sys0a} for models with parameters $\mathbf{P_{+}}$ and $\mathbf{P_{0}}$ and initial conditions $\mathbf{I_0}$, $\mathbf{I_1}$ and $\mathbf{I_2}$ .
\Fig{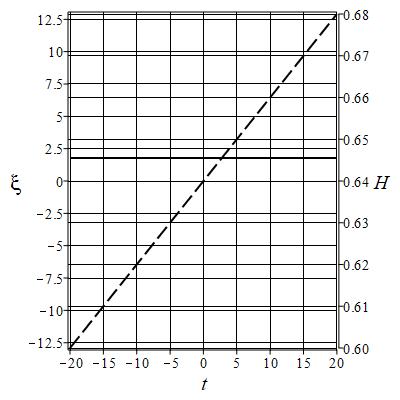}{8}{\label{Ign6}The scale function $\xi(t)$ is a dashed line and the Hubble parameter $H(t)$ is a solid line; $\mathbf{I}=\mathbf{I_0}$ (singular point $M_{0,-1}^{+}$).}

\Fig{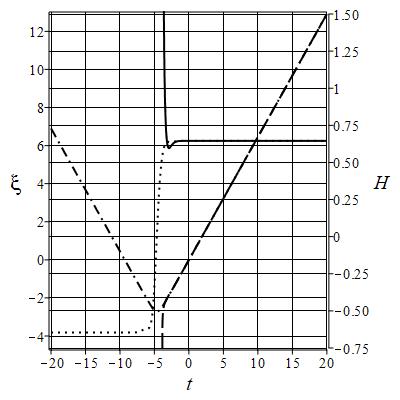}{8}{\label{Ign7}Scale function $\xi(t)$ and Hubble parameter $H(t)$. $\mathbf{P}=\mathbf{P_{+}}$: $\xi(t)$ -- dashed line, $H(t)$) -- solid; $\mathbf{P}=\mathbf{P_{0}}$: $\xi(t)$ -- dash-dotted, $H(t)$ -- dashed. Everywhere $\mathbf{I}=\mathbf{I_1}$.}

According to the graph of the function $\xi(t)$ in Fig.\ref{Ign8}, the moment of time $t_s\backsimeq-3.8539281$ corresponds to the cosmological singularity $a(t_s)\to 0$. Figure \ref{Ign9} shows a plot of the barotropic coefficient $\kappa(t)$ near this cosmological singularity. As can be seen from this graph, the barotropic coefficient tends to unity ($\kappa\to1$), both to the left and to the right of the singularity, in full accordance with the asymptotic estimate \eqref{varkappa=1}.

Commenting on the graphs in these figures, we note that the scale functions $\xi(t)$ and $H(t)$ of the models, corresponding to different values of the parameter $\gamma$, coincide at $t\gtrsim 0$, so the graphs at large times become indiscernible.

\FigH{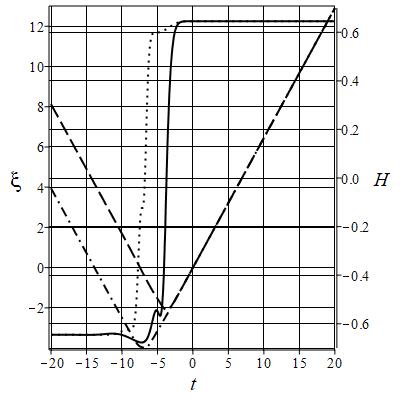}{8}{7}{\label{Ign8}Scale function $\xi(t)$ and Hubble parameter $H(t)$. $\mathbf{P}=\mathbf{P_{+}}$: $\xi(t)$ -- dashed line, $H(t)$) -- solid; $\mathbf{P}=\mathbf{P_{0}}$: $\xi(t)$ -- dash-dotted, $H(t)$ -- dashed. Everywhere $\mathbf{I}=\mathbf{I_2}$.}
\FigH{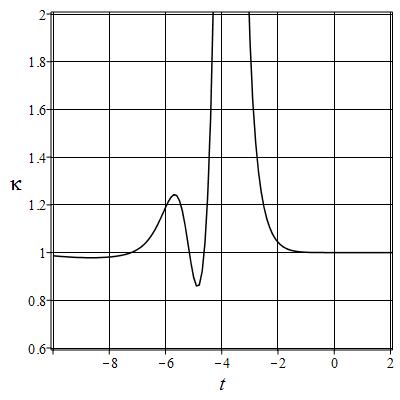}{7}{6}{\label{Ign9}Graph of the function $\kappa(t)$ in the model with parameters $\mathbf{P_{+}}$ under initial conditions $\mathbf{I_2}$.}

Note that the coordinates of a stable singular point are an exact stable solution of the dynamical system. To do this, consider the following initial conditions:
\begin{equation}\label{I_z}
\mathbf{I^\pm_z}=[0,0,0,1,\pm1].
\end{equation}
The graphs in Fig.\ref{Ign5} illustrate precisely this fact: trajectories corresponding to the initial conditions $\mathbf{I_1}$, $\mathbf{I_2}$ are attracted over time to this stable trajectory.
In Fig.\ref{Ign10} -- \ref{Ign11} shows graphs of three-dimensional projections $\Sigma_{\Phi}$ and $\Sigma_{\varphi}$ of the Einstein - Higgs hypersurface \eqref{Hip_Ein} cosmological model with parameters $\mathbf{ P_+}$ and initial conditions $\mathbf{I_0}$ at time $t=100$. Note that, generally speaking, the graphs of these projections depend on time, in contrast to the five-dimensional Einstein-Higgs hypersurface itself (for details, see \cite{TMF20}), however, if the initial conditions coincide with the coordinates of a special stable point, the projections remain invariant , their appearance for this case is shown in Fig.\ref{Ign10} -- \ref{Ign11}.

\vspace{-0.5cm}
\Fig{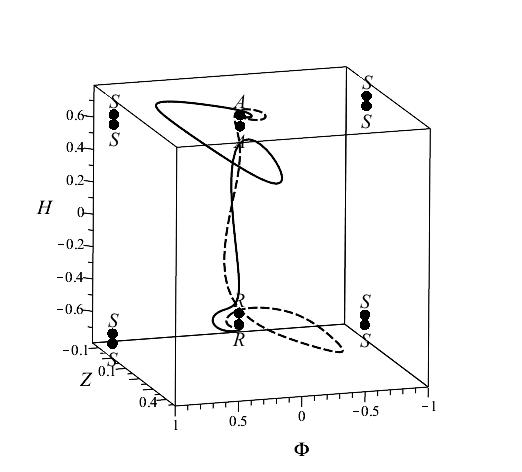}{9}{\label{Ign5}Phase trajectories of a classical scalar field: $\mathbf{P_+}$ \eqref{P_{+}}; $\mathbf{I^-_z}$ is a solid line and $\mathbf{I_z^{+}}=[0,0,0,1,1]$ is a dashed line.}
\Fig{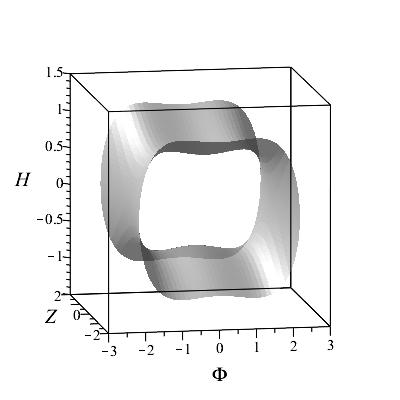}{8}{\label{Ign10}Projection of the Einstein-Higgs surface \eqref{Hip_Ein} onto the subspace $\Sigma_{\Phi}$}

In Fig.\ref{Ign12} -- \ref{Ign15} shows the evolution of projections when the initial conditions do not coincide with the coordinates of the special stable point $\mathbf{I_z}$. In this case, there is a significant change in the geometry and topology of the projections of the classical subspace $\Sigma_{\Phi}$, in contrast to the phantom $\Sigma_{\varphi}$, the geometry and topology of which do not undergo significant changes over time. Below everywhere: $\mathbf{P}=\mathbf{P_+}$ \eqref{P_{+}}, $\mathbf{I}=\mathbf{I_z}$ \eqref{I_z}. Dashed lines - $t=-10$, solid lines - $t=0$, long dashed lines - $t=1$, dotted lines - $t=2$, dash-dotted lines - $t=20$.

\Fig{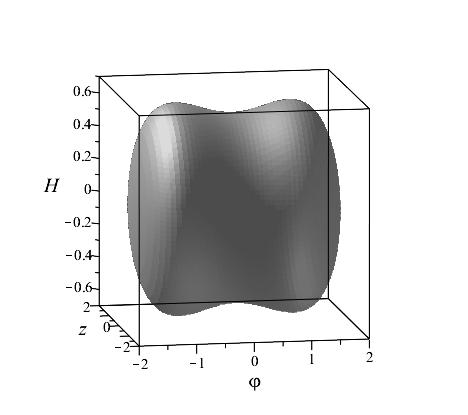}{8}{\label{Ign11}Projection of the Einstein-Higgs surface \eqref{Hip_Ein} onto the subspace $\Sigma_{\varphi}$.}

\subsubsection*{Evolution of the projection of the Einstein-Higgs surface \eqref{Hip_Ein} onto the subspace $\Sigma_{\Phi}$}

\FigH{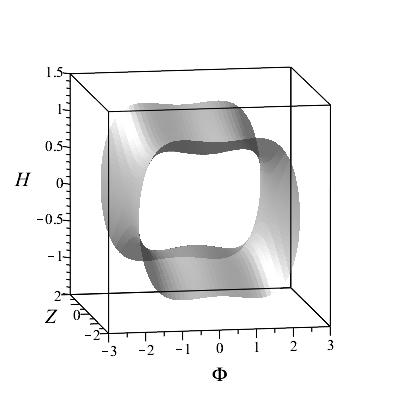}{7}{6}{\label{Ign12}$t=-10$}
\FigH{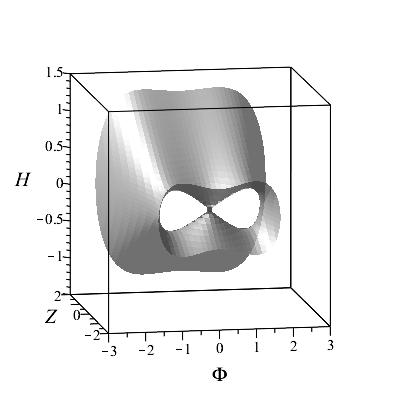}{7}{6}{\label{Ign13}$t=0$.}
\FigH{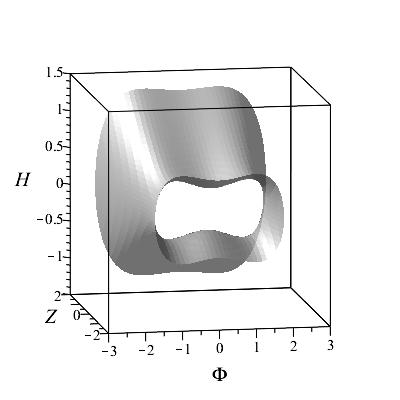}{7}{6}{\label{Ign14}$t=1$}
\FigH{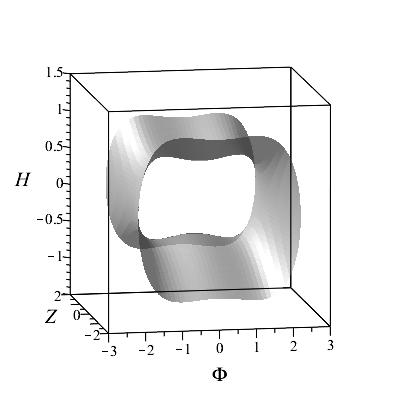}{7}{6}{\label{Ign15}$t=2$.}
\subsubsection*{Evolution of the section of projections of the Einstein - Higgs surface \eqref{Hip_Ein} onto the subspaces $\Sigma_{\Phi}$ and $\Sigma_{\varphi}$}
In Fig.\ref{Ign16} -- \ref{Ign17} shows graphs of the evolution of two-dimensional sections of the Einstein-Hip\-gs hypersurface \eqref{Hip_Ein}.
Everywhere: $\mathbf{P}=\mathbf{P_+}$ \eqref{P_{+}}; $\mathbf{I}=\mathbf{I^-_z}$ \eqref{I_z}; dashed lines - $t=-10$, solid - $t=0$, long dashed - $t=1$, dotted - $t=2$, dash-dotted - $t=20$.
\FigH{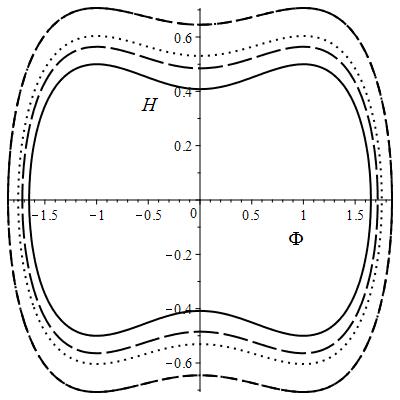}{6.5}{5.5}{\label{Ign16} Section of projections $Z=0$}
\FigH{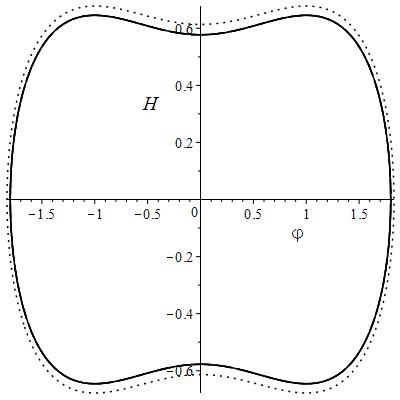}{6.5}{5.5}{\label{Ign17}Section of projections $z=0$.}

\section{Conclusion}
To summarize this part of the article, let us highlight the main ones:
\begin{itemize}
\item The simplest mathematical model of the evolution of the Universe is formulated, based on an asymmetric Higgs scalar doublet with a kinetic coupling proportional to the product of the derivatives of the components of the scalar doublet.
\item The symmetry properties of the model with respect to the reflection and similarity transformation have been studied and proven.
\item It is shown that in a number of cases the model has cosmological singularities at finite cosmological times, as in the case of models based on systems of scalarly charged fermions, but unlike the latter it has
asymptotic behavior near singularities, corresponding to an extremely rigid, rather than an ultra-relativistic, equation of state.
\item A qualitative analysis of the model under study was carried out.
\item Detailed numerical modeling of the constructed mathematical model was carried out for a simple set of fundamental constants and initial conditions that coincide with the parameters of the special stable point of the model and are close to them.
\end{itemize}
The next part of the article will be devoted to mathematical modeling and research of cosmological models in a wide range of fundamental constants and initial conditions and identifying the main types of behavior of models.

\subsection*{Founding}
The work is performed according to the Russian Government Program of Competitive Growth of Kazan Federal University.


\begin{thebibliography}{90}

\vspace{3mm}

\bibitem{TMF20} Yu. G. Ignat'ev, I. A. Kokh, Theoret. Math. Phys., \textbf{207:1}, 51 (2021); arXiv:2104.01054 [gr-qc].

%
\bibitem{Vernov1}
I.Ya. Aref’eva, A.S. Koshelev, and S.Yu. Vernov, Phys. Rev. D, \textbf{72:6}, 064017 (2005); arXiv: astro-ph/0507067.
%

%
\bibitem{Vernov2}
I. Ya. Aref’eva, S. Yu. Vernov, A. S. Koshelev, Theoret. Math. Phys., \textbf{148:1}, 895 (2006); arXiv:astro-ph/0412619.

\bibitem{Vernov3}
S. Yu. Vernov, Theoret. Math. Phys., \textbf{155:1}, 544 (2008); arXiv:astro-ph/0612487.
%

%
\bibitem{Leon18}
Genly Leon, Andronikos Paliathanasis, Jorge Luis Morales,  Eur. Phys. J. \textbf{C 78}, 753 (2018);  arXiv:1808.05634 [gr-qc].
%

\bibitem{Ignat_Sasha_Dima1}
Yu. G. Ignat'ev, A. A. Agathonov and D. Yu. Ignatyev,	Gravit. Cosmol., \textbf{27:4}, 338 (2021); arXiv:2203.11946 [gr-qc].

\bibitem{Ignat_Sasha_Dima2}
Yu. G. Ignat'ev, A. A. Agathonov and D. Yu. Ignatyev,	Gravit. Cosmol., \textbf{28:1}, 10 (2022); arXiv:2203.12766 [gr-qc]

\bibitem{TMF_21}
Yu. G. Ignat'ev and D.Yu. Ignatyev, Theoret. Math. Phys., \textbf{209:1}, 1437 (2021); arXiv:2111.00492 [gr-qc].

\bibitem{TMF_24}
Yu. G. Ignat’ev,	Theoret. Math. Phys., \textbf{219:1}, 688 (2024); arXiv:2307.13761 [gr-qc].

%
\bibitem{SMBH1e}
S. Gillessen, F. Eisenhauer, S. Trippe, T. Alexander, R. Genzel, F. Martins, T. Ott, Astrophys.J. \textbf{692} 1075 (2009);
arXiv:0810.4674 [astro-ph].

%
\bibitem{SMBH2e}
Sheperd Doeleman, Jonathan Weintroub, Alan E.E. Rogers  et al., Nature \textbf{455} 78 (2008); arXiv:0809.2442 [astro-ph].
%

\bibitem{Trakhtenbrote}
B.~Trakhtenbrot, Proc. Internat. Astronom. Union, \textbf{15(S356)}, 261 (2019); arXiv:2002.00972v2  [astro-ph.GA].

%
\bibitem{Zhue}
 Q. Zhu, Yu. Li, Yi. Li, M. Maji, H. Yajima, R. Schneider and  L. Hernquist, Notices of the Royal Astronomical Society. \textbf{514(4)}, 5583 (2022); arXiv:2012.01458v1  [astro-ph.GA].
%

%
\bibitem{Supermass_BH}
L. Arturo Urena-Lopez, Andrew R. Liddle, \emph{Phys. Rev}., \textbf{D66} (2002) 083005; arXiv:astro-ph/0207493.
%
\bibitem{Shadows}
Pedro V. P. Cunha, Carlos A. R. Herdeiro, Eugen Radu and Helgi F. R\'{u}narsson,  \emph{Internat. J. Mod. Phys. D}, \textbf{25:9} (2016) 1641021.
%
\bibitem{Soliton}
Philippe Brax, Jose A. R. Cembranos, Patrick Valageas, \emph{Phys. Rev}. D \textbf{101}, 023521 (2020); arXiv:1909.02614 [astro-ph.CO].
%

\bibitem{Ignatev_SBH}
Yu. G. Ignat'ev, Gravit. Cosmol., \textbf{27:1}, 30 (2021); \textbf{27:1}, 36 (2021);  \textbf{28:1}, 25 (2022); \textbf{28:3}, 275 (2022); \textbf{28:4}, 375 (2022); \textbf{29:2}, 163 (2023); \textbf{29:4}, 327 (2023);
\textbf{30:1}, 40 (2024); \textbf{30:2}, 141 (2024).


\bibitem{TMF_23}
Yu. G. Ignat'ev, Theoret. Math. Phys., \textbf{215:3} 862 (2023); arXiv:2306.17185 [gr-qc].

%
\bibitem{Land}
L. D. Landau, E. M. Lifshitz. \emph{The Classical Theory of Fields.} Pergamon Press. Oxford$\cdot$ New York$\cdot$ Toronto$\cdot$ Sydney$\cdot$ Paris$\cdot$ Frankfurt, 1971.
%


\bibitem{Ignat_Dima_GC20}
 Yu. G. Ignat’ev and D. Yu. Ignatyev, Grav. and Cosmol., {\bf 26}, 29  (2020);  arXiv:2005.14010 [gr-qc].


\bibitem{Similarity}
Yu. G. Ignat'ev, Theoret. Math. Phys., \textbf{219:1}, 688 (2024); arXiv:2307.13761v1 [gr-qc].

\bibitem{Bogoyav}
 O. I. Bogoyavlensky. \emph{The methods of the qualitative theory of dynamic systems in astrophysics and gas dynamics.} Nauka. Moskow, 1980.
%
%
\bibitem{Bautin}
N.N. Bautin, E.A. Leontocich. \emph{Methods and techniques for the qualitative study of dynamical systems on a plane}. Nauka. Moskow, 1989.
%
\end{thebibliography}
\end{document}